\documentclass[fleqn,usenatbib]{mnras}

\usepackage{newtxtext,newtxmath}

\usepackage[dvipsnames]{xcolor}
\usepackage[T1]{fontenc}
\usepackage{pdflscape}
\usepackage{multicol}

\DeclareRobustCommand{\VAN}[3]{#2}
\let\VANthebibliography\thebibliography
\def\thebibliography{\DeclareRobustCommand{\VAN}[3]{##3}\VANthebibliography}

\usepackage{graphicx}	
\usepackage{amsmath}

\newcommand{\JetSeT}{\texttt{JetSeT}}

\title[Jetted hybrid AGN sources]{Exploring the nature of the jetted hybrid AGNs: PKS 2004-447, 3C 286, and PKS 0440-00 through the SED modeling}

\author[J. Luna-Cervantes et al.]{
J. Luna-Cervantes,$^{1}$\thanks{E-mail: jluna@astro.unam.mx}
A. Tramacere,$^{2}$
E. Benítez$^{1}$\thanks{E-mail: erika@astro.unam.mx}
\\
$^{1}$Universidad Nacional Autónoma de México, Instituto de Astronomía, AP 70-264, CDMX 04510, Mexico\\
$^{2}$Department of Astronomy, University of Geneva, Ch. d’Ecogia 16, 1290 Versoix, Switzerland
}

\date{Accepted XXX. Received YYY; in original form ZZZ}

\pubyear{2024}

\begin{document}
\label{firstpage}
\pagerange{\pageref{firstpage}--\pageref{lastpage}}
\maketitle

\begin{abstract}

In this work, we explore the connection of three jetted $\gamma-$loud AGNs classes: Compact Steep-Spectrum Sources (CSS), Narrow-Line Seyfert 1 (NLS1), and Flat-Spectrum Radio Quasars, through the modeling of the spectral energy distribution (SED). We selected two sources identified as CSS/NLS1 hybrids, PKS\,2004-440 and 3C\,286. Additionally, we included the source PKS\,0440-00, initially classified as an FSRQ in the first \textit{Fermi}-LAT  catalog, but recently reclassified as an NLS1. We present the results of their broadband SED modeling using a one-zone leptonic synchrotron-self Compton (SSC)+ external Compton (EC) model. By exploring the parameter space and investigating the disk-jet connection in these sources, we analyze their classification in a model-dependent way.
Our findings reveal that modeling PKS\,2004-447 at relatively large angles, as expected for CSS, results in an SSC-dominated inverse Compton emission. In contrast, at low observing angles, the inverse Compton emission is dominated by external photon fields. Both scenarios result in a jet with a low radiative power.
For 3C\,286 we found that using a one-zone model limits the jet viewing angle to $\sim7^{\circ}$, mainly due to its impact on the $\gamma$-ray emission. Our model results show a magnetically dominated jet, consistent with $\gamma$-CSS sources.
Our results suggest that PKS\,0440-00, can be classified as a powerful $\gamma-$NLS1, characterized by high accretion power and a jet dominated by bulk motion, similar to FSRQs.

\end{abstract}


\section{Introduction}

The radio-loud Active Galactic Nuclei (AGNs), also known as jetted AGNs by \cite{2017NatAs...1E.194P}, are divided into different classes, including blazars and radio galaxies. The orientation of relativistic jets in these AGNs, relative to our line of sight, can lead to different observed characteristics \citep[][]{1993ARA&A..31..473A,1995PASP..107..803U}. For instance, their small viewing angles in blazars can result in extreme variability and high polarization emission. Conversely, in radio galaxies, the orientation enables the observation of extended radio lobes. Additional classes of jetted AGNs were also identified as gamma-ray sources by the Large Area Telescope (LAT) on board the \textit{Fermi} Gamma-Ray Space Telescope, in particular, some Narrow-line Seyfert 1 (NLS1) galaxies \citep[e.g.,][]{2009ApJ...699..976A} and Compact-steep Spectrum Sources (CSS) \citep{2020ApJS..247...33A}.

The formation of jets in AGNs is thought to be connected to a spinning, accreting supermassive black hole (SMBH), and the accretion disk formed by the falling material towards it. \citep[e.g.,][]{1971reas.book.....Z,1978Natur.275..516R,2019ARA&A..57..467B}. The intense magnetic fields near the black hole event horizon play a crucial role in the acceleration and collimation of particles into these relativistic jets \citep[e.g.,][]{1977MNRAS.179..433B,1984RvMP...56..255B,2013ApJ...764L..24S}.

Blazars are jetted AGNs known for their extreme and often variable emission across the electromagnetic spectrum, from radio to gamma-rays, harboring powerful and highly beamed relativistic jets \citep[e.g.][]{1978bllo.conf..328B,2007Ap&SS.309...95B}. Their multiwavelength spectral energy distribution (SED)  exhibits a characteristic double hump shape dominated by non-thermal emission processes originating from the relativistic jet. Based on the rest-frame equivalent width (EW) of their optical emission lines, blazars are commonly divided into Flat spectrum radio quasars (FSRQ) with EW$>$5\,\AA\, and the BL Lac objects (EW$<$\,5\,\AA). FSRQs are characterized by strong broad emission lines and a photon-rich environment, with a strong and efficient accretion disk \citep[][]{2012MNRAS.421.1569B}, while  BL\,Lac objects show weak or absent emission lines, low-density photon environment and a disk accreting inefficiently onto the central SMBH  \citep[][]{2012MNRAS.421.1764S}.

NLS1s are identified by their optical spectrum, which exhibits broad and narrow emission lines. The criteria commonly used to define NLS1s include showing narrow H$\beta$ emission lines with FWHM(H$\beta$) $<$\,2000\,km\,$s^{-1}$), weak [OIII] emission lines (with [OIII]/H$\beta<3$), and the presence of strong FeII multiplets \citep[][]{1985ApJ...297..166O,1989ApJ...342..224G}.

The SED of NLS1 galaxies often shows strong thermal emission from the accretion disk and non-thermal emission from the jet. Only $7\%$ of the NLS1 are radio-loud \citep[][]{2018rnls.confE..15K}, and within them, only a small fraction has been detected in gamma-rays.
This fraction of NLS1 shows steep photon indices and high variability in the X-ray band, which corroborates the relatively low-mass black holes (typically M$_{\mathrm{BH}}\lesssim$ 10$^8$\,M$_{\odot}$, in comparison with blazars, where M$_{\mathrm{BH}}\lesssim$ 10$^{8-10}$\,M$_{\odot}$), accreting at high Eddington rates \footnote{The Eddington ratio is defined as: $\lambda=L_{\mathrm{bol}}/L_{\mathrm{Edd}}$, where $L_{\mathrm{bol}}$ is the bolometric luminosity, and $L_{\mathrm{Edd}}=1.3\times 10^{38}M_{\mathrm{BH}}/M_{\odot}$.} \citep[][]{2018rnls.confE..15K}. Some authors have proposed that jetted NLS1s are the young counterpart of FSRQs and that those with misaligned relativistic jets may eventually evolve to form the parent population of FSRQs \citep[see e.g.,][]{2016A&A...591A..88B,2017A&A...603C...1F,2021A&A...654A.125B}.

About 70$\%$ of the NLS1 are compact radio-emitters with steep spectra, showing a connection with the Compact steep-spectrum sources (CSS) \citep[][]{2018rnls.confE..15K}. The link between NLS1 and CSS has been suggested by many authors \citep[e.g.,][]{2006AJ....132..531K,2015ApJS..221....3G,2016A&A...591A..88B}. The CSS are compact radio sources with a steep high-frequency spectrum in radio ($\alpha\geq0.5$, $F_{\nu}\propto\nu^{-\alpha}$) with a turnover below $\sim 400\,$MHz, characterized to have fully developed radio lobes and small projected linear size ($<20\,$kpc) which is considered to be an indicator of young age \citep[][]{1995A&A...302..317F,1998PASP..110..493O,1999A&A...345..769M}. Many CSS sources are embedded in a dense interstellar medium with their jets vigorously interacting with it \citep[see e.g.,][]{2021A&ARv..29....3O}.

The existence of sources that share properties associated with two classes (i.e., hybrids, in this case, NLS1s with blazar-like SEDs or radio characteristics similar to CSS), indicates a potential evolutionary connection between these jetted AGN classes, that could be understood in terms of a unification scenario \citep[][]{2021A&A...654A.125B}. The shared underlying physical processes offer a unique opportunity to explore the interplay between accretion disks, relativistic jets, and surrounding matter in AGNs.

In this work, we investigate the parameter space obtained through the SED modeling of three jetted AGNs gamma-ray emitters, using the \JetSeT\, framework, fitting the SED from radio to $\gamma$-rays with a one-zone leptonic SSC+EC model. We examine different models to fit the SEDs of the chosen sources.
This approach allows us to explore the nature of the sources by testing different orientations of the jet and diverse black hole masses to get the best fit of the SEDs of the chosen sources.
We then discuss the implications of the derived jet and disk properties, intending to identify signatures that associate the sources with a specific jetted AGN class or confirm their hybrid behavior. The selected sources include two objects previously classified as hybrid sources CSS/NLS1: PKS\,2004-447 and 3C\,286. We examine their similarities and differences with the NLS1 and CSS classes, exploring their classification.
Additionally, PKS\,0440-00 is incorporated into our analysis as a potential hybrid between the NLS1 and FSRQ classes. We present the best model for this source and discuss the feasibility of classifying it as NLS1 or proposing it as an FSRQ/NLS1 hybrid.

The paper is organized as follows: in Section \ref{sect:sample} we present the three selected sources. In Section \ref{sect:data}, we describe the multiwavelength data employed to construct the SED for each source. In section \ref{sect:model} we describe the leptonic model used to fit the broadband SED. The main results are presented in Section \ref{sect:res}. A detailed discussion of the nature of the three sources is given in Section \ref{sect:dis}, and we summarize our conclusions in Section \ref{sect:conc}. Throughout this paper we use a Hubble constant $H_0=67.8\,$km\,s$^{-1}$\,Mpc$^{-1}$, $\Omega_m=0.307$, and $\Omega_\varLambda=0.692$ \citep{2014A&A...571A..16P}.

\section{The Sample}
\label{sect:sample}

We have selected three $\gamma$-ray-emitting jetted AGNs, being classified either as Compact Steep Sources(CSS), Narrow Line Seyfert 1 (NLS1), or Flat Spectrum Radio Quasars (FSRQ). Two sources are already being classified as CSS/NLS1, i.e. hybrid sources, namely, PKS\,2004-447 ($z=0.24$) \citep[][]{2021A&A...654A.125B} and 3C\,286 ($z=0.849$) \citep[][]{2017FrASS...4....8B}. The hybrid nature was established based on their radio properties similar to CSS and optical emission lines in agreement with the NLS1 class. The third source, PKS\,0440-00 ($z=0.844$), initially detected in the $\gamma$-rays by the \textit{Fermi} satellite during its first year of operation, is known to be a bright FSRQ \citep[]{2010ApJ...716...30A}. However, \cite{2021Univ....7..372F} have found NLS1 characteristics on this source, in particular the FWHM(H$\beta$)=1700\,km\,s$^{-1}$, which associates PKS\,0440-00 with the NLS1 class, according to the FWHM(H$\beta$) < 2000\,km\,s$^{-1}$ criterion.

\section{Multiwavelength data}
\label{sect:data}

For each of the sources, we built the average SED, covering the radio band up to $\gamma$-rays, using data collected from the ASI Science Data Center (ASDC) SED builder\footnote{\url{https://tools.ssdc.asi.it/SED/}}.

The data collected for PKS\,2004-447 spans from August 2008 to May 2012, with $\gamma-$rays data from the \textit{Fermi}-LAT Third Source Catalog (3FGL). Based on the Fermi light curve data\footnote{Fermi LAT Light Curve Repository (LCR) \url{https://fermi.gsfc.nasa.gov/ssc/data/access/lat/LightCurveRepository/}}, the source did not exhibit significant activity during this period. The SED is representative of an average activity state.
For 3C\,286, we use simultaneous observations from the \textit{Swift} UVOT and XRT, taken in August 2020, analyzed by \cite{2021MNRAS.501.1384Y}. The SED covers observations from radio to $\gamma$-rays between August 2008 and August 2021, including data from the Atacama Large Millimeter/submillimeter Array (ALMA) Calibrator Source Catalogue\footnote{\url{https://almascience.eso.org/sc/}}. Despite the long time range, the source showed steady activity, particularly in $\gamma-$rays, with a variability index of 2.92, as reported in the fourth \textit{Fermi} Large Area Telescope catalog (4FGL).
For PKS\,0440-00, the collected data spans from March 2007 to February 2010, using $\gamma-$rays data from the 2 Year \textit{Fermi}-LAT Sources Catalog (2FGL). Even when the source had significant gamma-ray activity during the period covered by the SED, there were no critical flux changes according to the Fermi light curve. Moreover, we studied the average activity state of the source.

\section{modeling the broadband SED}
\label{sect:model}

We modeled the SED of the sources with a numerical one-zone leptonic model implemented in Jets SED modeler and fitting Tool (\texttt{JetSeT}) \citep{2009A&A...501..879T,2011ApJ...739...66T,2020ascl.soft09001T}. The model accounts for synchrotron radiation and inverse Compton (IC) scattering of both, synchrotron radiation (Synchrotron Self-Compton, SSC) and external photon fields (External Compton, EC), to fit the SED from radio to the $\gamma$-rays. The seed photons for the EC process are photons generated by the accretion disk surrounding the black hole (BH) and reflected towards the jet by the BLR and the dusty torus. The $\gamma$-ray emission from FSRQ, is well-fitted by EC processes \citep{2012ApJ...752L...4M,2013ApJ...763..134F}.

In the first stage, we preliminary constrain the parameter space, using the phenomenological constraining modules implemented in \texttt{JetSeT}. In the second stage, the minimization of the model was obtained using the \texttt{JetSeT} \texttt{ModelMinimizer} module, plugged to \texttt{iminuit} python interface \citep{dembinski_hans_2023_7750132}. Finally, we employ a Markov Chain Monte Carlo sampling, to estimate the posterior distribution of the physical parameters of the sources, and of the jet energetic, using the \texttt{JetSeT} \texttt{McmcSampler} module, plugged into the \texttt{emcee} sampler \citep{emcee}.

In the following section, we describe the details of the SSC/EC model and the relevance of the physical parameters.

\subsection{Model description}

The model assumes a homogeneous emission region with a spherical shape, located at a distance $R\mathrm{_{diss}}$ from the central BH of mass $M\mathrm{_{BH}}$. The emitting region travels along the jet axis with a bulk Lorentz factor $\Gamma$, with a viewing angle $\theta$, implying a beaming factor $\delta=1/(\Gamma(1 -\beta \cos\theta)$. Assuming that the jet has a conical shape, with a semi-opening angle represented by $\psi$, we introduce the relationship \footnote{Following the \href{https://jetset.readthedocs.io/en/latest/user_guide/documentation_notebooks/jet_model_phys_EC/Jet_example_phys_EC.html}{\JetSeT \,documentation} } between the size of the emitting region $R$ and $R\mathrm{_{diss}}$, expressing it as $R=R_{\mathrm{diss}}\mathrm{tan}~\psi$.   

Since cooling effects primarily influence the SED of the sources, we model them by assuming that the emitting region is filled with electrons following a broken power-law energy distribution, given by:

\begin{equation}
     N(\gamma)\propto  \left\lbrace
     \begin{array}{ll}
     \gamma^{-p} & \gamma\mathrm{_{min}}\leq \gamma < \gamma\mathrm{_{b}} \\
    \gamma^{-p_1} & \gamma\mathrm{_{b}}\leq \gamma\leq\gamma\mathrm{_{max}},
     \end{array}
     \right.
 \end{equation}

with the number density of the electrons
\begin{equation}
    N_e=\int_{{\gamma}_{\mathrm{{min}}}}^{\gamma\mathrm{_{max}}}N(\gamma)d\gamma,
\end{equation}

where, $\gamma\mathrm{_{min}}$, $\gamma\mathrm{_{b}}$ and $\gamma\mathrm{_{max}}$ are the minimum, break, and maximum electron Lorentz factor, respectively, and $p$ and $p_{1}$ the low-energy and high-energy spectral index. The relativistic electrons interact with the entangled magnetic field, $B$, emitting synchrotron and inverse Compton radiation (IC). For the IC, the full Compton cross-section for relativistic electrons including the Klein-Nishina regime is considered.

To reproduce the beaming pattern of the EC emission \citep{1995ApJ...446L..63D}, we use the \texttt{disk} external field transformation in \JetSeT\,, which follows the approach of \cite{2001ApJ...561..111G}, ensuring the correct dependence of the beaming pattern on the observing angle.

In cases where an optical-UV excess flux is observed in the SED, we can estimate the accretion disk physical parameters and set the black hole mass. On the other hand, in cases where the jet emission dominates, the values of $M_{\mathrm{BH}}$ are taken from the literature,  mainly derived from spectroscopy, through the luminosity of the broad emission lines, \citep[see e.g.,][]{2021A&A...654A.125B}.

We assume that the accretion disk extends from, $R_{\mathrm{in}}=3R_{\mathrm{S}}$, to $R_{\mathrm{out}}=500R_{\mathrm{S}}$ ($R_{\mathrm{S}}$ is the Schwarzschild radius, $R_{\mathrm{S}}=2GM_{\mathrm{BH}}/c^2$) with an emission profile that has a multi-temperature blackbody shape \citep[][]{2002apa..book.....F,2009MNRAS.397..985G}. The radial dependence of the temperature is given by:
\begin{equation}
     \mathrm{T(R)=\left\{\frac{3R_{\mathrm{S}}L_{\mathrm{Disk}}}{16\pi \eta\sigma R^3}\left[1-\left(\frac{R_{in}}{R}\right)^{1/2}\right]\right\}^{1/4}},
 \end{equation}
where $\sigma$ is the Stefan-Boltzmann constant, and $\eta$ is the accretion efficiency linked to the accretion rate $\dot{M}$, and to the bolometric luminosity. Assuming that $L_{\mathrm{bol}}\propto L_{\mathrm{Disk}}$, the disk luminosity is given by $L_{\mathrm{Disk}}=\eta \dot{M}c^2$. We have not considered the disk corona emission.

The BLR is assumed to be a spherical thin shell with a radius that scales with the disk luminosity following the relation\footnote{$L_{\mathrm{Disk,X}}$ is defined as the disk luminosity in 10$^{X}$\,erg\,s$^{-1}$ units.},
\begin{equation}
    R\mathrm{_{BLR}}=3\times10^{17}(L\mathrm{_{Disk,46}})^{1/2}\,\mathrm{cm},
\end{equation}
given by   \cite{2007ApJ...659..997K}. The DT is modeled as a black body with temperature $T_{\mathrm{DT}}$, constrained from $\sim$100\,K up to dust sublimation temperature $\sim$2000\,K \footnote{Sublimation temperature of graphite dust grains \citep{2017MNRAS.470.2578G}}
\citep[][]{2008MNRAS.387.1669G,2009MNRAS.397..985G}.  Its radius is estimated using the relation by \cite{2000ApJ...545..107B},
\begin{equation}
    R\mathrm{_{DT}}=2.5\times10^{18}(L\mathrm{_{Disk,45}})^{1/2}\,\mathrm{cm}.
\end{equation}
We assume a fraction $\tau\mathrm{_{BLR}}=0.1$ of $L\mathrm{_{Disk}}$ is reprocessed by the BLR, and a fraction $\tau\mathrm{_{DT}}=$0.1-0.3 by the DT \cite{2009MNRAS.397..985G}.
The effect of extragalactic background light absorption was taken into account \citep{2008A&A...487..837F}.

\subsubsection{Jet power}
We get the energetic content of the jet obtained from the SED modeling and the jet power carried by electrons ($P_{\mathrm{e}}$), Poynting flux ($P_\mathrm{B}$), radiation ($P_{\mathrm{rad}}$), and cold protons ($P_\mathrm{p}$) as follows \citep{2008MNRAS.385..283C}:
\begin{equation}
    P\mathrm{_i} =\pi R^2 \beta c\Gamma^2 U'_i,
    \label{power}
\end{equation}
where $R$ is the size of the emission region, $\beta c$ is the plasma velocity, and $U'_i$ is the comoving energy density of relativistic electrons ($i$=e), magnetic field ($i=$B), radiation ($i=$rad), and cold protons ($i=$p).
The single-sided jet power is given by:
\begin{equation}
    P_{\mathrm{jet}}=\pi R^2\beta c\Gamma^2(U'_\mathrm{e}+U'_\mathrm{B}+U'_\mathrm{p}+U'_\mathrm{r},
    ),
\end{equation}
and we have assumed one proton per electron to estimate the cold protons energy density.

The derived parameters of the modeling are displayed in Tables \ref{Tabl:params2004}, \ref{Tabl:params3c286}, and \ref{Tabl:params0440}. The best fits obtained by MCMC are shown as posterior contour maps for the relevant parameters of each model in Figures \ref{fig:2004-447_mcmc}, \ref{fig:3c286_mcmc_angle} and \ref{fig:0440_mcmc}, as well as a comparison of the derived jet power posterior distribution with a sample of $\gamma-$NLS1 and FSRQ in Figures \ref{fig:2004-447_power}, \ref{fig:3c286_power} and \ref{fig:posterior_0440}.

\section{Results}
\label{sect:res}

We have modeled the average SED of the three sources to derive the physical parameters associated with the accretion disk, the jet, and the surrounding photon environment. In this section, we present the results derived from the fit of the SED using \JetSeT.

\subsection{PKS 2004-447}
\label{sect:2004res}

To determine whether the physical characteristics of the $\gamma$-ray emitter PKS\,2004-447 confirm its hybrid nature (NLS1/CSS), we examined the following models. Motivated by the hypothesis that NLS1 galaxies can be considered as CSS viewed at small angles, we model the SED for different viewing angles \citep[e.g.][]{2001ApJ...558..578O,2006AJ....132..531K,2016A&A...591A..88B}.

Previous studies modeled the SED of PKS\,2004-447 assuming the jet is oriented very close to the line of sight, as expected for $\gamma-$NLS1 sources \citep[$\theta=2^{\circ}$ and 3$^{\circ}$, see e.g.,][]{2013ApJ...768...52P,2021A&A...649A..77G}.
Also, different authors have found that the jet of PKS\,2004-447 may have a non-negligible inclination, similar to what is expected for CSS. Particularly, \cite{2016A&A...588A.146S} estimate an upper limit of  $\theta<50^{\circ}$ based on radio maps. Therefore, in addition to modeling the SED with a small viewing angle of $\theta=2^{\circ}$, we have investigated a scenario where the jet is oriented at a larger viewing angle, assuming an upper limit around $\theta=15^{\circ}$ \citep[larger angles result in a notable reduction in gamma-ray emission, attributed to the angle dependence in the Compton-scattering processes, see e.g.,][]{2001ApJ...561..111G,2016ApJ...830...94F}.

The best model obtained to fit the multi-wavelength data collected for PKS\,2004-447 is shown in Figure \ref{fig:2004-447_MBerton}. As we can observe, the low-energy hump is dominated by the jet emission, and the big blue bump produced by the accretion disk is not visible. In this case, to estimate $L_{\mathrm{Disk}}$, we constrained this parameter within a range going from $(1\times10^{42}-4\times10^{44})\mathrm{\,erg\,s^{-1}}$, based on the values reported by different authors on this source \citep[][]{2015MNRAS.453.4037O,2017A&A...603C...1F,2021A&A...654A.125B}. Since in the literature, low $M_{\mathrm{BH}}$ values have been used for modeling the broadband SED of PKS\,2004-447, we have taken a different approach using a larger value of the $M_{\mathrm{BH}}$, i.e. $M\mathrm{_{BH}}=6\times\,10^8\,M_{\odot}$ estimated by \citet{2016MNRAS.458L..69B}. Additionally, we examined a case with a lower value derived from the optical spectrum, $M\mathrm{_{BH}}=1.5\times 10^7\,M_{\odot}$ \citep[][]{2021A&A...654A.125B}. For each BH mass value, we present two models: one scenario with $\theta=2$ and another testing a larger viewing angle.

To explore the size of the blob and its distance from the BH, we have set the opening angle of the jet as a free parameter, constrained by a maximum value of 5.72$^{\circ}$, as is usually assumed for blazars \citep{2009MNRAS.397..985G}. The opening angle and the blob size derived from the fit, allow us to calculate the Doppler factor $\delta$ and variability timescale $\Delta \tau_v$ reported in Table \ref{Tabl:params2004}. In addition, we have computed the Eddington luminosity and the Eddington ratio. The parameters obtained with the best-fit models are presented in Table \ref{Tabl:params2004}.

\subsubsection{Models assuming high black hole mass}

Within the framework of the jet oriented at $\theta=2^{\circ}$, adopting a $M\mathrm{_{BH}}=6\times 10^8\,M_{\odot}$ \cite{2016MNRAS.458L..69B}, we have derived a $L_{\mathrm{Disk}}=1.33\times 10^{43}\,\mathrm{erg\,s^{-1}}$ (it is important to be cautious with this luminosity value obtained from the fit, since the optical/UV spectrum does not allow direct observation of the accretion disk emission). In this model, the first hump of the SED from IR to UV data is well reproduced assuming synchrotron emission, and the second hump at high energies is mainly dominated by the EC process generated by the dusty torus photon field. We derived a magnetic field strength of $B=0.53\,$G, and a Bulk Lorentz factor of $\Gamma=10.07$.
The emitting region is located outside the BLR but inside the dusty torus.

In the second scenario, we tested a jet oriented with a viewing angle larger than 2 degrees, and $\theta$ set as a free parameter constrained by an upper limit of $\theta=15^{\circ}$. We also set as free parameters the magnetic field and the parameters associated with the electron distribution to fit the SED. We obtained from the best fit a $\theta=9.5^{\circ}$, with a statistic $\chi_r^2=1.99$. In contrast to the model with $\theta=2^{\circ}$, the X-ray and $\gamma-$ray data are explained by SSC emission, and the low-frequency radio data are well reproduced by the synchrotron emission. We derived a low Bulk Lorentz factor $\Gamma=4.37$, and a low magnetic field $B=0.018\,$G, required to reproduce the $\gamma-$ray emission. The size of the emitting region is estimated of $R_{\mathrm{blob}}=4.77\times 10^{17}\,\mathrm{cm}$, located at a distance $R_{\mathrm{diss}}=5.77\times 10^{18}\,\mathrm{cm}$ from the BH, beyond the DT radius.

\subsubsection{Models assuming low black hole mass}

When we assume a black hole mass of $M\mathrm{_{BH}}=1.5\times 10^7\,M_{\odot}$ \citep[][]{2021A&A...654A.125B}, we find that, at an angle of $\theta=2^{\circ}$, the parameters derived from the fit are very similar to those obtained with the model assuming a BH mass of $M\mathrm{_{BH}}=6\times 10^8\,M_{\odot}$ \citep[][]{2016MNRAS.458L..69B}, as we can observe in the Table \ref{Tabl:params2004}. The low-frequency hump is well explained by synchrotron radiation, and the gamma-ray emission is effectively reproduced by EC emission dominated by seed photons from the dusty torus. Similarly, when we set the viewing angle as a free parameter, the derived parameters resemble those obtained from the model assuming the higher mass; reproducing the SED with an SSC-dominated model but, this time obtaining a $\theta=10.8^{\circ}$. The main difference with the models where we assumed a higher black hole mass, lies in the estimated Eddington rate, which depends on the black hole mass and disk luminosity.

\begin{figure*}
	\includegraphics[width=\textwidth]{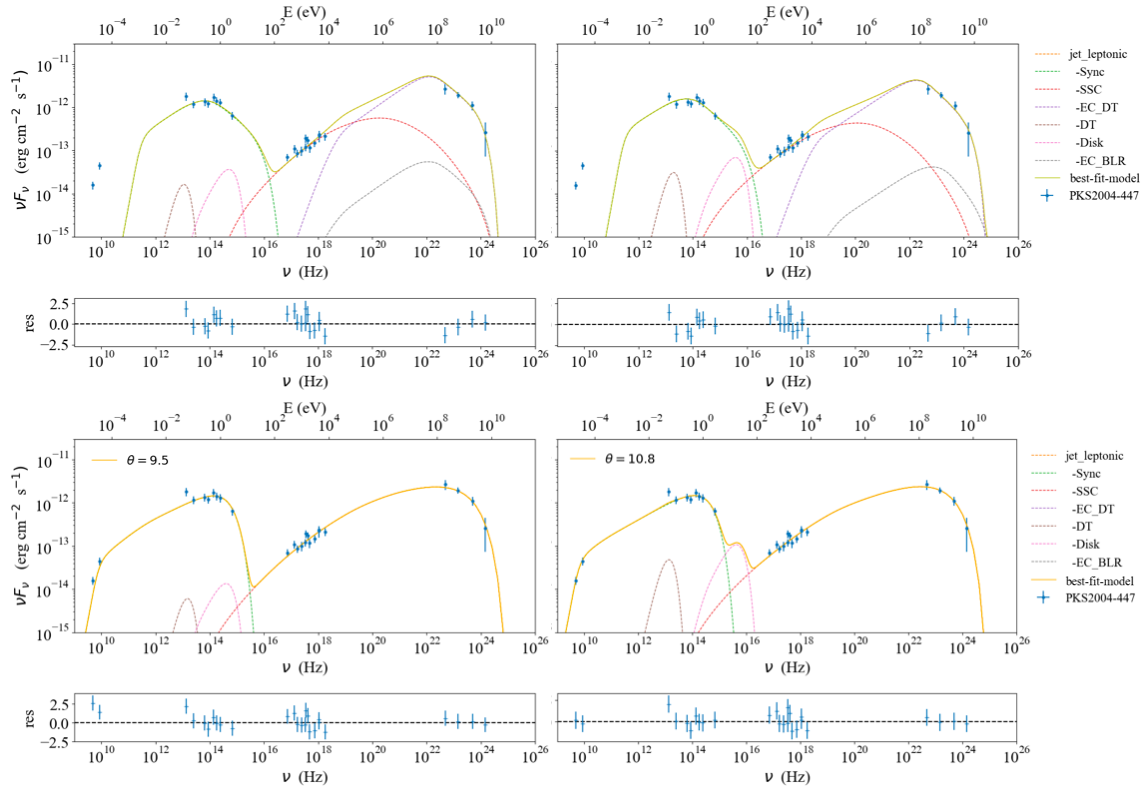}
    \caption{Top left panel: Broadband SED of PKS\,2004-447 modeled assuming a viewing angle of $\theta=2^{\circ}$ and a $M\mathrm{_{BH}}\sim6\times 10^8\,M_{\odot}$ \citep{2016MNRAS.458L..69B}. The low-energy hump is reproduced by synchrotron emission, the dusty torus component with a peak at IR frequencies, and the accretion disk component. The high-energy hump is reproduced by a combination of the SSC emission and EC scattering of the BLR and dusty torus photons. The best fit was achieved with a model where the low-energy bump is dominated by synchrotron emission, while the high-energy bump is reproduced by SSC+EC emission, with external Compton dominated by photons coming from the dusty torus.
    Bottom left panel: The best fit assuming a viewing angle larger than $\theta=2^{\circ}$ and the same $M\mathrm{_{BH}}$ as in the top panel. From the fit, a jet oriented at $\theta=9.5^{\circ}$ was derived. In this case, the high-energy bump is dominated by SSC emission. Top right panel: The best-fit model for the SED of PKS\,2004-447 assuming a $M\mathrm{_{BH}}\sim1.5\times 10^7\,M_{\odot}$ \citep{2021A&A...654A.125B}, and a viewing angle of $\theta=2^{\circ}$. The low-energy hump is dominated by synchrotron emission, and the high-energy hump is dominated by EC scattering of dusty torus photons. Bottom right panel: Model assuming the viewing angle as a free parameter, obtaining $\theta=10.8^{\circ}$ from the fit. The high-energy hump is dominated by SSC emission.}
    \label{fig:2004-447_MBerton}
\end{figure*}

\begin{table}
 \caption{ Column (1): Parameter name. Column (2): Model of PKS\,2004-447 assuming a viewing angle of $\theta=2^{\circ}$ and $M_{\mathrm{BH}}$ taken from \citep{2016MNRAS.458L..69B}. Column (3): Same as Column 2, but with a larger viewing angle. Column (4): Model of PKS\,2004-447 assuming a viewing angle of $\theta=2^{\circ}$ and $M_{\mathrm{BH}}$ taken from \citep{2021A&A...654A.125B}. Column (5): Same as Column 4, but with a larger viewing angle. }
 \label{Tabl:params2004}
 \begin{tabular*}{\columnwidth}{l@{\hspace*{12pt}}l@{\hspace*{12pt}}l@{\hspace*{12pt}}l@{\hspace*{12pt}}l@{\hspace*{12pt}}} \hline
Parameter                                     & \multicolumn{4}{c}{PKS\,2004-447} \\ \hline
(1)                                           & (2)       & (3)    & (4)  & (5)  \\ \hline
p                                             & 2.10      & 1.96    & 2.16     & 1.96    \\
p$_{1}$                                       & 3.98      & 2.38    & 4.12      & 2.29  \\
$\gamma'\mathrm{_{min}}$                      & 26.51     & 53.74  & 15.57     & 60.13  \\
$\gamma'\mathrm{_{b}}$                        & 1.5$\times10^{3}$  & 1.27$\times10^{3}$     & 1.42$\times10^{3}$  & 526.4    \\
$\gamma'\mathrm{_{max}}$                      & 1.32$\times10^4$    & 3.81$\times10^{4}$  & 1.33$\times10^4$    & 4.01$\times10^{4}$  \\
$N\,[\mathrm{cm^{-3}}]$                       & 4.76$\times10^{3}$  & 8.97   & 7.04$\times10^{3}$  & 5.74   \\
$B\,\mathrm{[G]}$                             & 0.53     & 0.018   & 0.62     & 0.014    \\
$\Gamma$                                      & 10.07    & 4.37    & 10.0    & 4.20    \\
$R\mathrm{_{diss}}$\,[cm]                     & 4.73$\times10^{16}$  & 5.77$\times10^{18}$  & 5.73$\times10^{16}$  & 9.75$\times10^{18}$  \\
$R\mathrm{_{blob}}$\,[cm]                     & 3.44$\times10^{15}$ & 4.77$\times10^{17}$  & 3.82$\times10^{15}$ & 7.09$\times10^{17}$  \\
$\Psi$\,[deg]                                 & 4.12     & 4.69     & 3.37     & 4.12  \\
$\theta$ [deg]                                & 2.0$^f$  & 9.5    & 2.0$^f$  & 10.8\\
$M\mathrm{_{BH}\,[M_{\odot}]}$                & 6$\times10^{8^f}$   & 6$\times10^{8^f}$   & 1.5$\times10^{7^f}$   & 1.5$\times10^{7^f}$  \\
$L\mathrm{_{Disk}}$[{$\mathrm{erg\,s^{-1}}$}] & 1.33$\times10^{43}$    & 5$\times10^{42}$    & 2.53$\times10^{43}$    & 4$\times10^{43}$ \\
$\eta$                                        & 0.08$^f$            & 0.08$^f$            & 0.08$^f$                 & 0.08$^f$     \\
$R\mathrm{_{DT}}^f$[cm]                       & 2.89$\times10^{17}$ & 1.76$\times10^{17}$ & 3.97$\times10^{17}$ & 5$\times10^{17}$  \\
$T\mathrm{_{DT}}$[K]                          & 170      & 230      & 299      & 208   \\
$\tau\mathrm{_{DT}}$                         & 0.3$^f$  & 0.3$^f$   & 0.3$^f$ & 0.3$^f$  \\
$\tau\mathrm{_{BLR}}$                         & 0.1$^f$  & 0.1$^f$   & 0.1$^f$ & 0.1$^f$  \\
$R\mathrm{_{BLR,in}}^f$[cm]                   & 1.15$\times10^{16}$ & 7.07$\times10^{15}$  & 1.59$\times10^{16}$ & 2$\times10^{16}$   \\
$R\mathrm{_{BLR,out}}^f$[cm]                  & 1.27$\times10^{16}$ & 7.77$\times10^{15}$  & 1.75$\times10^{16}$ & 2.2$\times10^{16}$ \\ \hline
$P\mathrm{_{e}}$\,[$\mathrm{erg\,s^{-1}}$]    & 43.67               & 44.92     & 43.68               & 45.04   \\
$P\mathrm{_{B}}$\,[$\mathrm{erg\,s^{-1}}$]    & 42.11               & 42.71     & 42.32               & 42.84    \\
$P\mathrm{_{r}}$\,[$\mathrm{erg\,s^{-1}}$]    & 42.12               & 43.34     & 42.08               & 43.47    \\
$P\mathrm{_{p}}$\,[$\mathrm{erg\,s^{-1}}$]    & 44.90               & 45.73     & 45.16               & 45.84    \\
$P\mathrm{_{tot}}$\,[$\mathrm{erg\,s^{-1}}$]  & 44.93               & 45.79     & 45.17               & 45.91    \\ \hline
$U'\mathrm{_e}$\,[$\mathrm{erg\,s^{-3}}$]     & 4.13$\times10^{-1}$ & 2.11$\times10^{-3}$  & 3.56$\times10^{-1}$ & 1.36$\times10^{-3}$   \\
$U'\mathrm{_B}$\,[$\mathrm{erg\,s^{-3}}$]     & 1.15$\times10^{-2}$ & 1.29$\times10^{-5}$  & 1.53$\times10^{-2}$ & 8.61$\times10^{-6}$ \\ \hline
$L\mathrm{_{Edd}}$[$\mathrm{erg\,s^{-1}}$]     & 7.8$\times10^{46}$ & 7.8$\times10^{46}$ & 1.95$\times10^{45}$ & 1.95$\times10^{45}$  \\
$\lambda\mathrm{_{Edd}}$                      & 1.7$\times10^{-4}$  & 6.4$\times10^{-5}$   & 0.012  & 0.02 \\
$\delta$                                      &   17.89   &   4.63    &   17.79   &   4.89   \\
$\Delta \tau_v$ [days]                        &   0.09     &  49.31   &   0.10     &  69.44  \\
$\chi^2_\mathrm{r}$/d.o.f                     &   1.82/12  &  1.99/13 &   1.71/12  &  1.19/13 \\ \hline
\hline
\multicolumn{5}{l}{$^f$ Fixed parameters during the fit}\\
\end{tabular*}
\end{table}

\begin{figure}
	\includegraphics[width=\columnwidth]{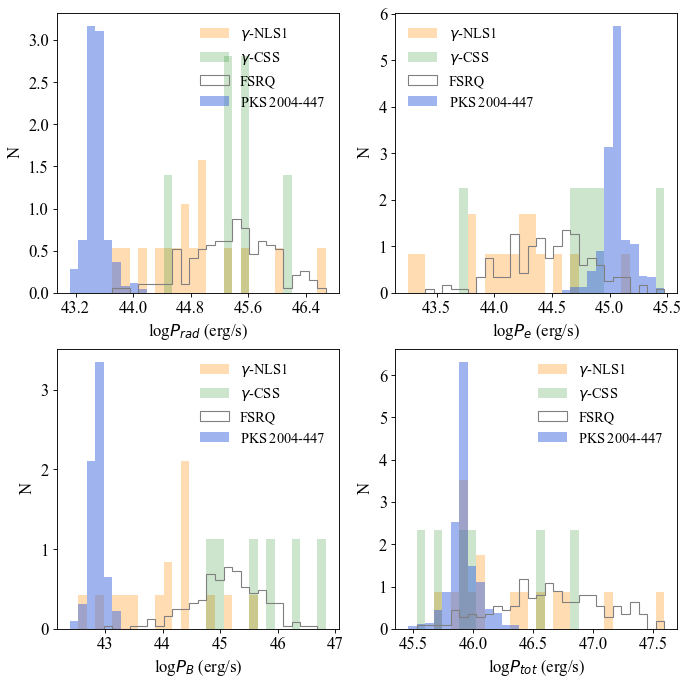}
    \caption{Posterior distribution of jet power components of PKS\,2004-447, obtained from the model presented in Column (5) of Table \ref{Tabl:params2004}, including radiation, electrons, and magnetic content, derived through the MCMC technique. The figure illustrates as a comparison, the jet power components of a sample of FSRQs (grey hatched area), $\gamma$-NLS1s (orange shaded area) and $\gamma$-CSSs (green shaded area), taken from \citet{2014Natur.515..376G,Paliya_2019,2020ApJ...899....2Z}, respectively. FSRQs predominantly exhibit a jet dominated by $P_\mathrm{rad}$, where $P_\mathrm{p}>P_\mathrm{rad}$. This translates to matter dominated jets. $\gamma$-CSS jets, as characterized by \citet{2020ApJ...899....2Z}, display high magnetization and radiation efficiency. $\gamma$-NLS1, share a composition similar to that of blazars, with the power of the particles dominating the total jet power.}
    \label{fig:2004-447_power}
\end{figure}

\subsection{3C 286}
\label{sect:3Cres}

To explore the hybrid nature of 3C\,286 (CSS/NLS1), we present two models of the SED. These models consider the jet oriented at two different angles. Before conducting the broadband modeling, we determined the black hole mass by fitting the simultaneous optical-UV data using the \JetSeT\, model for the accretion disk, deriving a black hole mass of $M_{\mathrm{BH}}=3.5\times10^8\,\mathrm{M_{\odot}}$. We then assumed this value in both models. We have built the radio to $\gamma$-rays SED, using the simultaneous optical-UV-X-ray data measured on August 2020, by the \textit{Swift} telescope and published by \cite{2021MNRAS.501.1384Y}, in addition to the archival data collected from the ASDC SED Builder. The collected SED along with the best fit associated with the models are shown in Figure\ref{fig:minuit3c286}. The parameters derived from the SED fit are reported in Table \ref{Tabl:params3c286}.

\subsubsection{Model assuming a small viewing angle}

We assumed an observing angle of $\theta=3^{\circ}$ as usually adopted for $\gamma$-NLS1s \citep[see e.g.][]{2009ApJ...707L.142A,Paliya_2019,2021A&A...649A..77G}, we fixed the black hole mass and set the disk luminosity and accretion efficiency as free parameters. Given the strong emission observed in the SED in the optical/UV bands coming from the accretion disk, we have constrained the range of $\eta$ to 0.06-0.1 \citep[see e.g.,][]{2015MNRAS.448.1060G}. This range is consistent with the radiative efficiency of the accretion disk. From the fit, we obtained a disk luminosity of $L_{\mathrm{Disk}}=6.6\times10^{46}\,\mathrm{erg\,s^{-1}}$, and an accretion efficiency of $\eta=0.072$.
The best fit is shown in the top panel of Figure \ref{fig:minuit3c286}, where the low-energy hump, is well reproduced by a combination of synchrotron emission, thermal emission from the dusty torus, and a significant contribution from the accretion disk. For the high-energy hump, we accurately reproduced the X-ray emission with a combination of the emission of the synchrotron high-energy tail, the SSC, and the EC emission from the dusty torus. The best fit for the $\gamma$-rays came mainly from the EC driven by seed photons from the BLR, with a contribution from the dusty torus seed photons. The dissipation region is located at $R_{\mathrm{diss}}=5\times10^{18}\,$cm, outside the BLR but within the dusty torus radius. From the obtained parameters, the Eddington ratio $\lambda_{\mathrm{Edd}}=1.52$ and the Doppler factor is $\delta=16.80$.

\subsubsection{Misaligned jet model}

Assuming that the jet is oriented at a larger viewing angle, consistent with the observed jets in the CSS class, we set $\theta$ as a free parameter constrained from 3$^{\circ}$ to 15$^{\circ}$. We used the parameters derived from the model with a small viewing angle as a starting point for the fit.  The parameters associated with the accretion disk component: $L_{\mathrm{Disk}}$, $M_{\mathrm{BH}}$, and the accretion efficiency, were fixed. Meanwhile, we set the magnetic field, the bulk Lorentz factor, the size of the emitting region, and its distance along the jet as free parameters.

\begin{figure*}
	\includegraphics[width=0.8\textwidth]{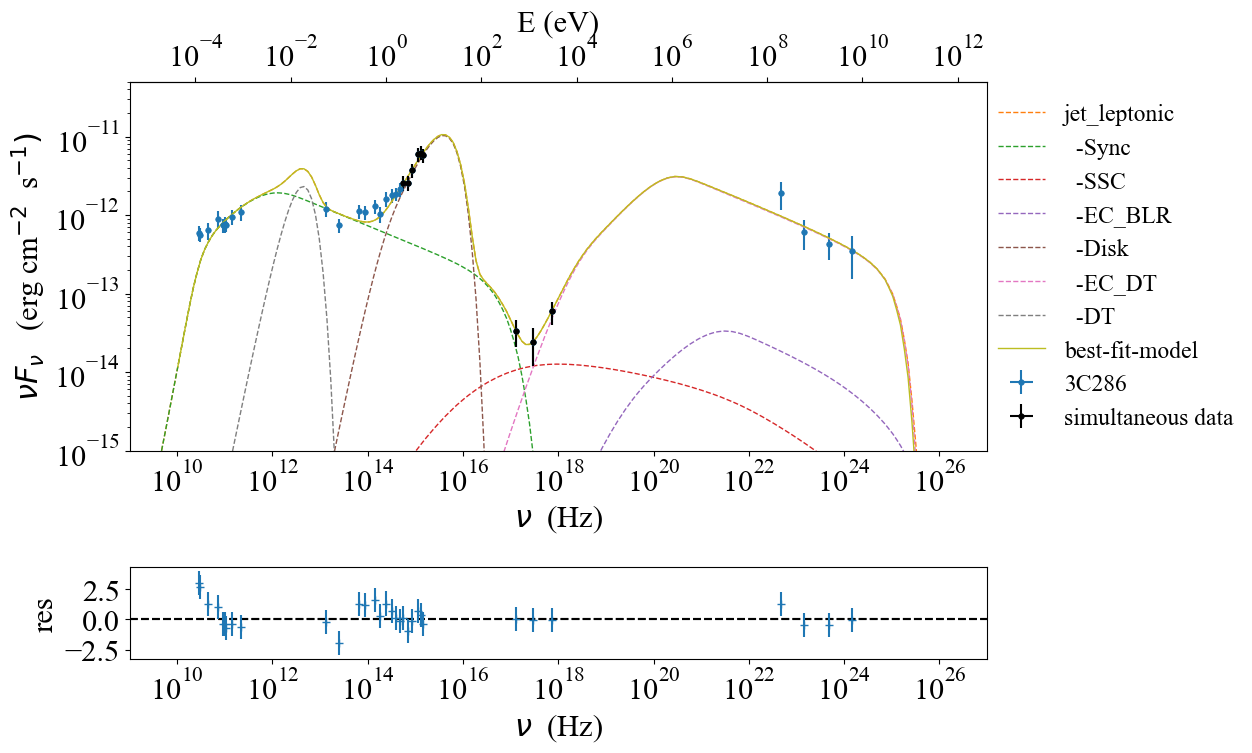}\\
    \includegraphics[width=0.8\textwidth]{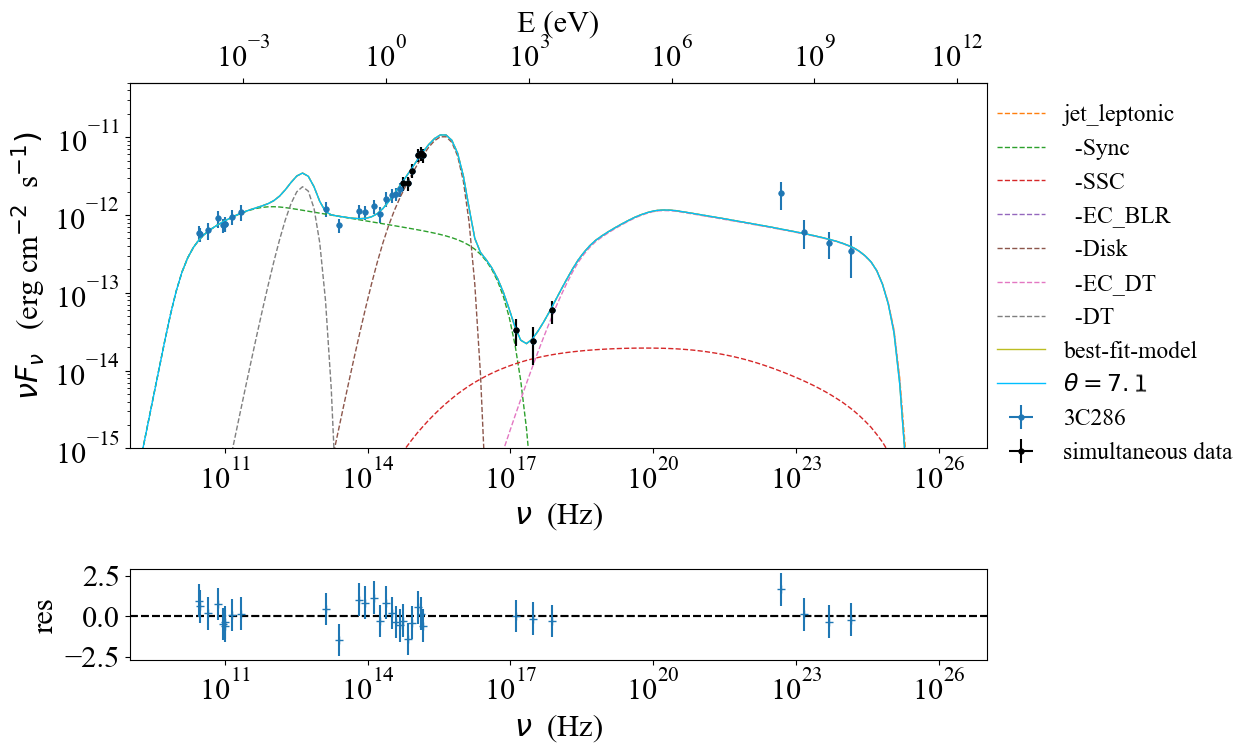}
    \caption{Top panel: Broadband SED and model fit of 3C\,286 assuming a viewing angle of $\theta=3^{\circ}$. The low-energy hump is reproduced by synchrotron emission, a dusty torus component peaking at IR frequencies, and the accretion disk component. The high-energy hump is reproduced by a combination of the SSC emission and EC scattering of the BLR and dusty torus photons.
    Bottom panel: The best model fit for the CSS scenario; with a derived observing angle of $\theta\approx7.1^{\circ}$. The high-energy hump is dominated by the torus-EC emission.}
    \label{fig:minuit3c286}
\end{figure*}

Due to the dependence of EC emission on $\theta$ \citep[see e.g.,][]{1995ApJ...446L..63D}, we found that it is not possible to fit the SED using a viewing angle larger than $\sim8^{\circ}$. As a result, our best fit was found with $\theta=7.1\pm0.3$. The model converges to a dissipation region located at $R_{\mathrm{diss}}=2.22\times10^{19}\,\mathrm{cm}\approx7.19\,$pc, an emitting region size $R_{\mathrm{blob}}=1.77\times10^{18}\mathrm{cm}$, and a magnetic field of $B=0.14\,$G. As observed in Figure \ref{fig:minuit3c286}, in this model, the self-absorption frequency, $\nu_{\rm SSA}\propto \Big(B^{\frac{p+2}{2}}RN\Big)^{\frac{2}{p+4}}$  \citep[see e.g.,][]{2013LNP...873.....G}, is redder than in the  model with $\theta=3^{\circ}$, allowing us to fit the radio data. The $\gamma$-rays are well reproduced with EC emission dominated by seed photons from the dusty torus.

\begin{table}
 \caption{ Column (1): Parameter name. Column (2): Parameters obtained from the SED fit of 3C\,286, assuming a viewing angle of 3$^{\circ}$. Column (3): Model of 3C\,286 for the CSS scenario.}
 \label{Tabl:params3c286}
 \begin{tabular*}{0.85\columnwidth}{l@{\hspace*{30pt}}l@{\hspace*{30pt}}l@{\hspace*{30pt}}l}
  \hline
Parameter          & \multicolumn{2}{c}{3C\,286}                             \\ \hline
(1)                                           & (2)                 & (3)               \\ \hline
p                                            & 1.67                & 2.2                \\
p$_{1}$                                      & 3.5                & 3.2                 \\
$\gamma'\mathrm{_{min}}$                     & 30.0                & 67.1                    \\
$\gamma'\mathrm{_{b}}$                        & 316.2                 & 456.7                   \\
$\gamma'\mathrm{_{max}}$                      & 8.0$\times10^4$       & 1.18$\times10^{5}$  \\
$N\,[\mathrm{cm^{-3}}]$                        & 1.03               & 0.18                \\
$B\,\mathrm{[G]}$                             & 0.22                & 0.14               \\
$\Gamma$                                      & 11.39                & 8.31             \\
$R\mathrm{_{diss}}$\,[cm]                     & 5$\times10^{18}$  & 2.22$\times10^{19}$  \\
$R\mathrm{_{blob}}$\,[cm]                     & 4$\times10^{17}$ & 1.77$\times10^{18}$ \\
$\Psi$\,[deg]                                 & 4.58$^f$            & 4.58$^f$           \\
$\theta$ [deg]                                & 3.0$^f$             & 7.1               \\
$M\mathrm{_{BH}\,[M_{\odot}]}$                & 3.5$\times10^{8^f}$  & 3.5$\times10^{8^f}$    \\
$L\mathrm{_{Disk}}$[{$\mathrm{erg\,s^{-1}}$}] & 6.6$\times10^{46}$  & 6.6$\times10^{46}$$^f$   \\
$\eta$                                        & 0.072            & 0.069       \\
$R\mathrm{_{DT}}^f$[cm]                       & 5.26$\times10^{19}$ & 5.13$\times10^{19}$  \\
$T\mathrm{_{DT}}$[K]                          & 100                 & 100                  \\
$\tau\mathrm{_{DT}}$                         & 0.15$^f$  & 0.15$^f$    \\
$\tau\mathrm{_{BLR}}$                         & 0.1$^f$  & 0.1$^f$   \\
$R\mathrm{_{BLR,in}}^m$[cm]                   & 7.89$\times10^{17}$    & 7.7$\times10^{17}$     \\
$R\mathrm{_{BLR,out}}^m$[cm]                  & 8.68$\times10^{17}$  & 8.47$\times10^{17}$   \\ \hline
$P\mathrm{_{e}}$\,[$\mathrm{erg\,s^{-1}}$]    & 44.28               & 44.74         \\
$P\mathrm{_{B}}$\,[$\mathrm{erg\,s^{-1}}$]    & 45.58               & 46.22         \\
$P\mathrm{_{r}}$\,[$\mathrm{erg\,s^{-1}}$]    & 43.69               & 44.71         \\
$P\mathrm{_{p}}$\,[$\mathrm{erg\,s^{-1}}$]    & 45.48               & 45.75         \\
$P\mathrm{_{tot}}$\,[$\mathrm{erg\,s^{-1}}$]  & 45.85               & 46.37         \\ \hline
$U'\mathrm{_e}$\,[$\mathrm{erg\,s^{-3}}$]     & 9.9$\times10^{-5}$  & 2.7$\times10^{-5}$     \\
$U'\mathrm{_B}$\,[$\mathrm{erg\,s^{-3}}$]     & 1.95$\times10^{-3}$  & 8.29$\times10^{-4}$     \\ \hline
$L\mathrm{_{Edd}}$[{$\mathrm{erg\,s^{-1}}$}]     & 4.55$\times10^{46}$ & 4.55$\times10^{46}$  \\
$\lambda\mathrm{_{Edd}}$    & 1.52 & 1.45 \\
$\delta$ &   16.80   &   8.09    \\
$\Delta \tau_v$  [days]     &      17.00      &      156.81         \\
$\chi^2_\mathrm{r}$/d.o.f                     & 1.81/20             &     0.66/23       \\ \hline
\hline
\multicolumn{3}{l}{$^f$ Fixed parameters during the fit}\\
\multicolumn{3}{l}{$^m$ model dependent parameters. $R\mathrm{_{BLR,out}}$ is fixed at $1.1R\mathrm{_{BLR,in}}$}\\
 \end{tabular*}
\end{table}

\begin{figure}
	\includegraphics[width=\columnwidth]{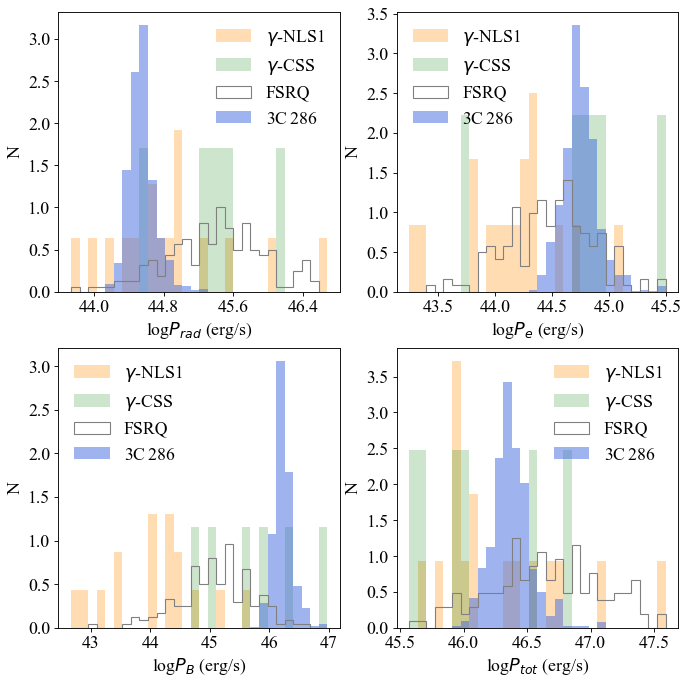}
    \caption{ Same as Figure \ref{fig:2004-447_power}, but for the jet power components of 3C\,286, obtained from the model presented in Column (3) of Table \ref{Tabl:params3c286}.}
    \label{fig:3c286_power}
\end{figure}

\subsection{PKS 0440-00}
\label{sect:pks0040res}

We have modeled the SED of PKS\,0440-00 using parameters typical of FSRQs, in agreement with its initial classification. Two models are presented: one constraining the emission region to be within the BLR and another placing it outside the BLR but within the dusty torus. In both models, we adopt a small viewing angle of $\theta=3^{\circ}$ and a jet opening angle of $\Psi=5.72^{\circ}$ \citep[see e.g.,][]{2015MNRAS.448.1060G}. By fitting the optical/UV data, we derived the black hole mass and accretion disk luminosity, obtaining a \textbf{$M\mathrm{_{BH}}= 1.73\times 10^{8} \mathrm{M_{\odot}}$} and $L\mathrm{_{Disk}}= 1.71\times10^{46} \mathrm{erg\,s^{-1}}$. These values were fixed in both models, assuming a radiative efficiency of $\eta=0.08$ \citep[see e.g.,][]{2009MNRAS.397..985G}. To estimate the radius of the dusty torus component, we extended the functional dependency based on the disk luminosity, and discussed in section \ref{sect:model}, with the sublimation temperature and the size of the dust grains \citep{2007A&A...476..713K}, according to:

\begin{equation}
\label{Rdt}
    R_{\mathrm{DT}}= 1.3 \left( \frac{L_{\mathrm{UV}}}{10^{46} \mathrm{erg/s}} \right) ^{1/2}\left( \frac{T_{\mathrm{sub}}}{1500\mathrm{K}}\right)^{-2.8}\left(\frac{a}{0.05\mu \mathrm{m}}\right)^{-1/2}\mathrm{\,cm},
\end{equation}
where we assumed that the dust concentration at the radius $R_{\mathrm{DT}}$, is mainly dominated by graphite grains with a size $a=0.05\,\mu m$ at $T_{\mathrm{sub}}$, which is considered as a free parameter within the range $T_{\mathrm{sub}}=[1000-2000]\,$K \citep[see e.g.,][]{2000ApJ...545..107B,2007A&A...476..713K}. $L_{\mathrm{UV}}$, is defined as the UV-optical luminosity of the central engine, which we have assumed as $L_{\mathrm{Disk}}$. The collected multiwavelength data, along with the best fit models are shown in Figure \ref{fig:0440_minuit}.  The
parameters obtained with the best-fit models are presented in Table \ref{Tabl:params0440}.

\subsubsection{Model with the emission region located outside the BLR but inside the dusty torus}
From the best fit we have found that the emission region is located at a distance of $R_{\mathrm{diss}}=1.4\times 10^{18}\,\mathrm{cm}$ from the BH. In this model (see top panel of Figure \ref{fig:0440_minuit}), we successfully reproduce the low energy hump of the SED with a combination of synchrotron emission and the thermal components from the dusty torus and accretion disk. We found a temperature of the DT of $\sim1693\,$K. The high energy hump is well reproduced with an SSC+EC emission, where the main contribution to the X-ray emission comes from the SSC. The EC used to fit the gamma-rays is dominated by photons from the DT.

\subsubsection{Model with the emission region within the BLR radius}
\begin{figure*}
    \centering
	\includegraphics[width=0.8\textwidth]{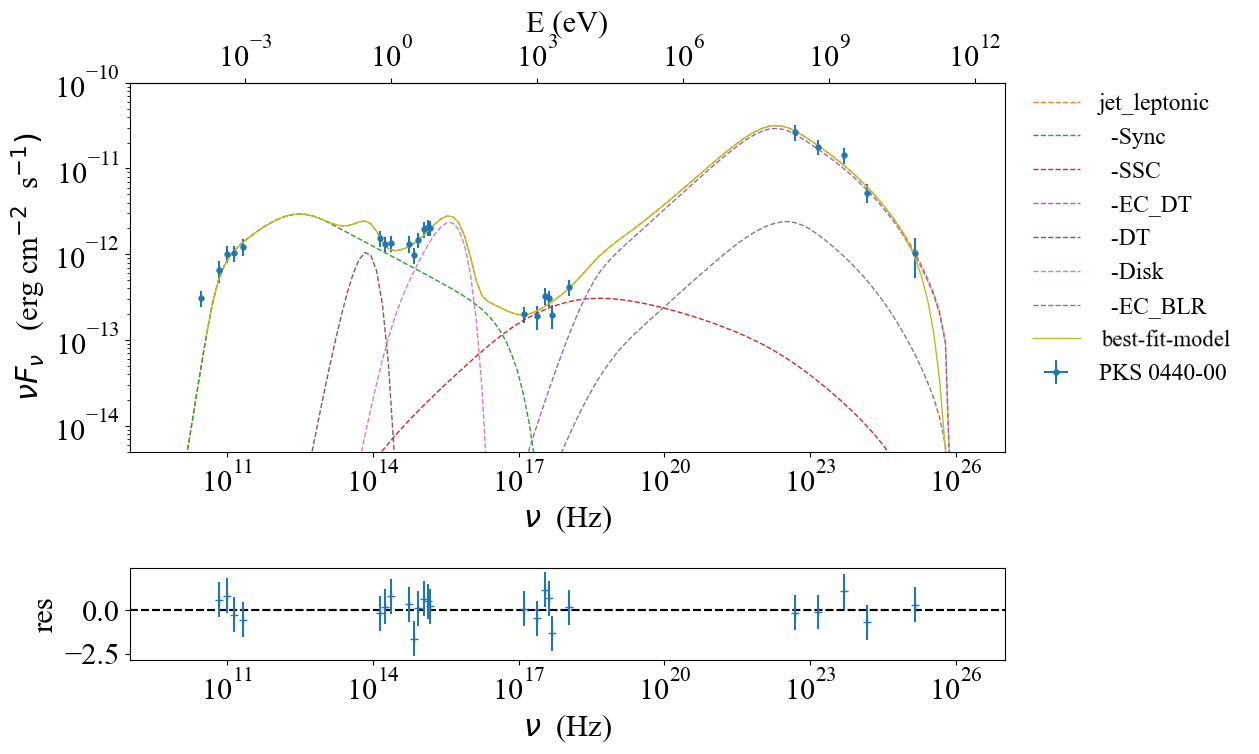}\\
    \includegraphics[width=0.8\textwidth]{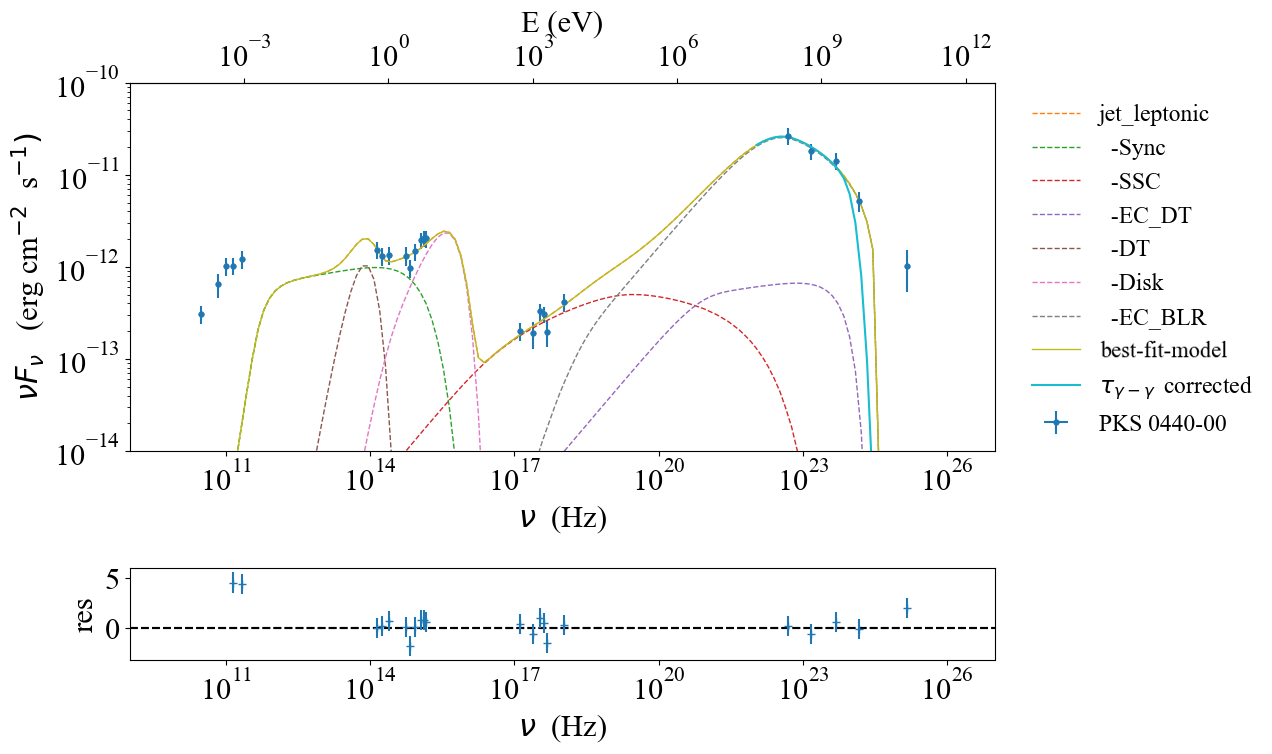}
    \caption{Top panel: Modeled broadband SED of PKS\,0440-00 assuming a dusty torus radius scaling following equation \ref{Rdt}. Bottom panel: Model of the SED of PKS\,0440-00 assuming the dissipation region located inside the BLR. In blue the $\gamma$-ray emission is corrected by the $\gamma-\gamma$ absorption.}
    \label{fig:0440_minuit}
\end{figure*}

When we assumed that the emission region is located inside the BLR, we found that the model fails to reproduce the radio emission (see bottom panel in Figure \ref{fig:0440_minuit}), and the gamma-rays are noticeably absorbed around 10 GeV energies. To obtain a better fit, we have calculated the optical depth for $\gamma$-ray photons and corrected the observed flux for $\gamma-\gamma$ absorption. In this model, the dominant component driving gamma-ray emission is the EC produced by seed photons from the BLR. The X-rays are well fitted with an SSC component.

\begin{table}
 \caption{Column (1): Parameter symbol. Column (2): Parameters obtained from the SED fit of PKS\,0440-00, taking into account the sublimation radius of the dust, assuming graphite grains with a size of $0.05\,\mu m$, for a $T_{\mathrm{DT}}=[1000,2000]\,$K. Column (3): Testing the dissipation region closer to the BLR radius.}
 \label{Tabl:params0440}
 \begin{tabular*}{0.8\columnwidth}{l@{\hspace*{20pt}}l@{\hspace*{20pt}}l@{\hspace*{20pt}}l@{\hspace*{20pt}}l}
  \hline
\hline
Parameter                     & \multicolumn{2}{c}{PKS\,0440-00}                       \\ \hline
(1)                                           & (2)                 & (3)    \\ \hline
p                                             & 1.96                & 1.77     \\
p$_{1}$                                       & 3.60                & 2.84     \\
$\gamma'\mathrm{_{min}}$                      & 7.69                   &  1$^f$  \\
$\gamma'\mathrm{_{b}}$                        & 627.2  & 200.0  \\
$\gamma'\mathrm{_{max}}$                      & 8.8$\times10^{4}$  & 4.1$\times10^3$  \\
$N\,[\mathrm{cm^{-3}}]$                       & 146.83 & 1.18$\times10^{5}$   \\
$B\,\mathrm{[G]}$                             & 0.15                & 1.8     \\
$\Gamma$                                      & 15.49               & 11.21   \\
$R\mathrm{_{diss}}$\,[cm]                     & 1.4$\times10^{18}$ & 1$\times10^{17}$   \\
$R\mathrm{_{blob}}$\,[cm]                     & 1.4$\times10^{17}$  & 5$\times10^{15}$   \\
$\Psi$\,[deg]                                 & 5.72$^f$                & 2.86$^f$  \\
$\theta$ [deg]                                & 3.0$^f$                & 3.0$^f$  \\
$M\mathrm{_{BH}\,[M_{\odot}]}$                & 1.73$\times10^{8^f}$   & 1.73$\times10^{8^f}$  \\
$L\mathrm{_{Disk}}$[{$\mathrm{erg\,s^{-1}}$}] & 1.71$\times10^{46^f}$  & 1.71$\times10^{46^f}$   \\
$\eta$                                        & 0.08$^f$                & 0.08$^f$ \\
$R\mathrm{_{DT}}^f$[cm]                       & 2.18$\times10^{19}$ & 1.79$\times10^{19}$ \\
$T\mathrm{_{DT}}$[K]                          & 1693               & 1800   \\
$T\mathrm{_{sub}}$[K]                         &  900  &  965 \\
$\tau\mathrm{_{DT}}$                          & 0.3$^f$   & 0.3$^f$  \\
$\tau\mathrm{_{BLR}}$                         & 0.1$^f$   & 0.1$^f$  \\
$R\mathrm{_{BLR,in}}^m$[cm]                   & 3.93$\times10^{17}$ & 3.93$\times10^{17}$  \\
$R\mathrm{_{BLR,out}}^m$[cm]                  & 4.32$\times10^{17}$ & 4.32$\times10^{17}$ \\\hline
$P\mathrm{_{e}}$\,[$\mathrm{erg\,s^{-1}}$]    & 45.33               & 44.44  \\
$P\mathrm{_{B}}$\,[$\mathrm{erg\,s^{-1}}$]    & 44.62              & 43.59  \\
$P\mathrm{_{r}}$\,[$\mathrm{erg\,s^{-1}}$]    & 44.55               & 44.36  \\
$P\mathrm{_{p}}$\,[$\mathrm{erg\,s^{-1}}$]    & 46.98               & 46.72   \\
$P\mathrm{_{tot}}$\,[$\mathrm{erg\,s^{-1}}$]  & 47.00              & 46.73  \\ \hline
$U'\mathrm{_e}$\,[$\mathrm{erg\,s^{-3}}$]     & 4.89$\times10^{-3}$ & 9.34$\times10^{-1}$  \\
$U'\mathrm{_B}$\,[$\mathrm{erg\,s^{-3}}$]     & 9.60$\times10^{-3}$ & 1.3$\times10^{-1}$ \\ \hline
$L\mathrm{_{Edd}}$[{$\mathrm{erg\,s^{-1}}$}]     & 2.25$\times10^{46}$ & 2.25$\times10^{46}$  \\
$\lambda\mathrm{_{Edd}}$    & 0.76 & 0.76 \\
$\delta$ &   18.69   &   16.66 \\
$\Delta \tau_v$ [days]    &      5.33      &      0.21 \\
$\chi^2_\mathrm{r}$/d.o.f      &       0.86/13        &    4.56/12 \\ \hline
\hline
\multicolumn{2}{l}{$^f$ Fixed parameters during the fit}\\
\multicolumn{3}{l}{$^m$ model dependent parameters. $R\mathrm{_{BLR,out}}$ is fixed at $1.1R\mathrm{_{BLR,in}}$}\\
 \end{tabular*}
\end{table}

\begin{figure}
    \includegraphics[width=\columnwidth]{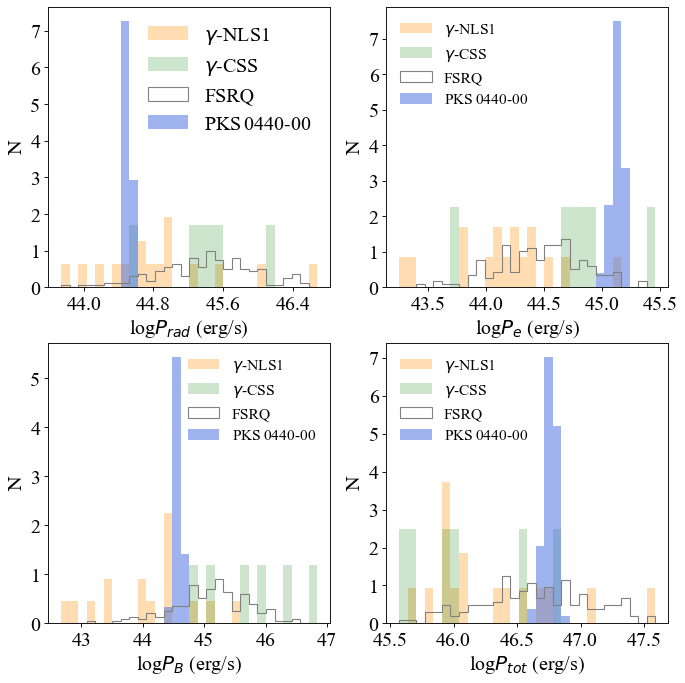}
    \caption{Same as Figure \ref{fig:2004-447_power}, but for the jet power components of PKS\,0440-00, obtained from the model presented in Column (2) of Table \ref{Tabl:params0440}.}
    \label{fig:posterior_0440}
\end{figure}

\section{Discussion}
\label{sect:dis}
In this section, we will discuss the physical parameters derived from the models presented in section \ref{sect:res}, to understand the hybrid nature of the sources.

\subsection{PKS 2004-447}

The multiwavelength SED of this source, built with the collected data indicates a blazar-like nature, with the gamma-ray emission confirming the presence of a relativistic jet. In the optical bands, this source is classified as an NLS1, due to its relatively weak [OIII] emission and narrow H$\beta$ line \citep{2001ApJ...558..578O,2021A&A...654A.125B}. In radio, characteristics such as the steep spectral index \citep[$\alpha < -0.5$, see e.g.,][]{2006MNRAS.370..245G} and core-jet morphology suggest a CSS classification \citep[see e.g.,][]{2001ApJ...558..578O,2006MNRAS.370..245G,2016A&A...588A.146S}. In addition, flux density measurements display a moderate slow variability in radio ($\sim20\%$ on the timescale of years) that are commonly found in CSS sources \citep[][]{2004A&A...424...91E}.

Our best model shows that the low-energy hump is dominated by jet emission, i.e. synchrotron radiation. This is also the case when modeling different activity states, a result recently reported by \cite{2021A&A...649A..77G}. On the other hand, the high-energy hump, as seen in Figure \ref{fig:2004-447_MBerton}, can be reproduced by either an SSC-dominated model or a combination of SSC+EC processes, depending on the orientation of the jet.

We obtain from the best fit of all the models, a relatively low $L\mathrm{_{Disk}}\sim10^{43}\mathrm{erg\,s^{-1}}$ that is in agreement with recent work by \cite{2021A&A...649A..77G}. Since the SED is dominated by jet emission, it is not possible to disentangle the contribution of the disk from that of the jet, therefore we might be underestimating the true luminosity of the disk. However, we found that values close to or above $\sim10^{44}\mathrm{erg\,s^{-1}}$, overestimate the SED flux in the optical/UV bands.

\subsubsection{The jet orientation}

Adopting a small viewing angle ($\theta=2^{\circ}$), our best fit was obtained with an SSC+EC model, where the gamma-rays are dominated by the EC emission produced by the DT (see top panels of Figure \ref{fig:2004-447_MBerton}). Similar results have been found also by other studies \citep[see e.g.,][]{Paliya_2019,2021A&A...649A..77G}. On the other hand, when modeling the SED with a free viewing angle, yielding an angle around 10$^{\circ}$, both X-rays and gamma-rays are well reproduced by SSC emission (see bottom panel of Figure \ref{fig:2004-447_MBerton}).
Given the observed short timescale variations in gamma-rays from hours to weeks, as noted by \cite{2021A&A...649A..77G}, the smaller viewing angle appears to be a more reliable estimate for this object. However, radio band observations suggest the possibility of a larger angle ($\theta\lesssim50^{\circ}$), as reported by \cite{2016A&A...588A.146S} based on their radio maps. Therefore, our estimated viewing angle of $\sim10^{\circ}$ remains valid (it is worth mentioning that it is important the scale at which this angle is estimated since the jet can undergo orientation changes). Moreover, this value is consistent with those found for $\gamma-$CSS sources \citep[][]{2020ApJ...899....2Z}.

\subsubsection{Black hole mass and accretion regime}

For modeling the SED, two different black hole masses were assumed. From \cite{2021A&A...654A.125B}, a black hole mass of $1.5\times10^7\,M_{\odot}$ was considered since its estimation seems to be less contaminated by the jet emission. On the other hand, we also used a second value of $6\times10^8\,M_{\odot}$ estimated by \cite{2016MNRAS.458L..69B}. These authors have proposed a higher estimated value, which may be more reliable. They suggest that the BLR clouds could be constrained within a rotating disk, potentially leading to a narrower FWHM of the spectral lines. As a result, this could lead to lower estimated black hole masses.

Different black hole masses have an impact on the accretion rate in Eddington units ($\lambda_\mathrm{Edd}=L_\mathrm{bol}/L_\mathrm{Edd}$). We have found that depending on the adopted BH mass and the $L_\mathrm{bol}$, the accretion could change from an efficient to an inefficient regime (see Table \ref{Tabl:accretion2004}).

To establish the accretion regime for the four models, we first assume $L_\mathrm{Disk}$ as a proxy of $L_\mathrm{bol}$. With this approximation, we have found that the models assuming $6\times10^8\,M_{\odot}$ indicate an inefficient accretion regime \citep[the transition between efficient to inefficient accretion mode occurs when $\lambda_{\mathrm{Edd,Disk}}\sim0.01$,][]{2010ASPC..427..249G}, while assuming $1.5\times10^7\,M_{\odot}$ suggests an efficient mode. A second approach was done considering now the $L_\mathrm{BLR}$, as a tracer of accretion \citep[see e.g.,][]{2014MNRAS.445...81S}. 

When considering $L_\mathrm{BLR}$, as a proxy of $L_\mathrm{Disk}$, based on the broad emission lines, the transition between efficient and inefficient regime is defined $\lambda_{\mathrm{Edd,BLR}}\sim 5\times 10^{-4}$\citep[][]{2014MNRAS.445...81S}. Under the accretion falls in the inefficient regime when we adopt a BH mass of $M\mathrm{_{BH}}=6\times 10^8\,M_{\odot}$, while with $M\mathrm{_{BH}}=1.5\times 10^7\,M_{\odot}$, the accretion disk becomes efficient. Given that the SED of this source is dominated by jet emission, with no prominent emission from the accretion disk, it is likely that considering $L_\mathrm{Disk}$ as an approximation of $L_\mathrm{bol}$ may not be the most appropriate assumption for this source.

Typically, the accretion disk in jetted NLS1 sources is considered to be efficient \citep[see e.g.,][]{2009ApJ...707L.142A}. Thus, assuming a mass of $1.5\times 10^7\,M_{\odot}$, the source seems more consistent with the NLS1 class. However, assuming a BH mass of $M\mathrm{_{BH}}=6\times 10^8\,M_{\odot}$, we found that the disk is accreting inefficiently. In this case, the accretion will be better described by an advection-dominated accretion flow model \citep[ADAF, see e.g.,][and references therein]{1998tbha.conf..148N}. According to \cite{2022MNRAS.517.1381C}, an ADAF-type model is the most suitable for this source, providing support for this particular scenario. Having a low accretion rate and a radiatively inefficient disk would naturally explain the relatively low luminosity of the disk.

Low Eddington ratio sources are characterized by weak or no Fe\,II emissions, as is the case of this source \citep[see e.g.,][]{2021A&A...654A.125B}. However, it does exhibit intense oxygen and Balmer emission lines in the optical, characteristic of the NLS1 class, indicating that the accretion disk is producing enough ionizing luminosity. As found by \cite{2011MNRAS.414.2674G}, even BL\,Lacs, that are low accretion sources can show strong broad emission lines depending on the jet activity state \citep[][]{2014MNRAS.445...81S}. The shape of the SED indicates that the jet dominates the emission, hiding the thermal emission from the accretion disk, a characteristic of the BL\,Lac class. It would be important to consider that this object could be in a transition phase where the accretion is low and the emission lines are the remnants of a previous more active state, i.e. an efficient accretion disk. Based on the above results of the SED modeling, we could support the hybrid nature of the source.

\begin{table}
 \caption{Parameters calculated from the models of PKS\,2004-447, with the corresponding black hole mass and viewing angle. $R_{\mathrm{BLR}}$ is assumed as $R_{\mathrm{BLR,out}}$, derived from the fit and reported in Table \ref{Tabl:params2004}. $L_{\mathrm{BLR}}=4\pi cR^2_{\mathrm{BLR}}U_{\mathrm{BLR}}$ where $U_{\mathrm{BLR}}$ is measured at the disk rest frame, provided by \JetSeT. $\dot{m}$ was estimated as $\dot{m}=\dot{M}/\dot{M_{\mathrm{Edd}}}$, where $\dot{M}=\frac{L_{\mathrm{Disk}}}{\eta c^2}$ and $\dot{M_{\mathrm{Edd}}}=\frac{L_{\mathrm{Edd}}}{c^2}$.}
 \label{Tabl:accretion2004}
 \begin{tabular*}{\columnwidth}{l@{\hspace*{7pt}}l@{\hspace*{9pt}}l@{\hspace*{9pt}}|l@{\hspace*{9pt}}l@{\hspace*{9pt}}}
\hline \hline
 & \multicolumn{2}{c|}{$M_{\mathrm{BH}}=6\times 10^8\,M_{\odot}$} &  \multicolumn{2}{c}{$M_{\mathrm{BH}}=1.5\times 10^7\,M_{\odot}$}   \\ \hline
 &  $\theta=2^{\circ}$ & $\theta=9.5^{\circ}$  & $\theta=2^{\circ}$ & $\theta=10.8^{\circ}$ \\
(1) & (2) & (3) & (4) & (5)\\ \hline
$R_{\mathrm{BLR}}$\,[cm]            & 1.27$\times10^{16}$ & 7.77$\times10^{15}$ & 1.75$\times10^{16}$ & 2.19$\times10^{16}$\\
$U_{\mathrm{BLR}}$\,[erg\,cm$^{-3}$]    & 2.66$\times10^{-2}$   & 3.13$\times10^{-2}$ & 2.55$\times10^{-2}$  & 2.48$\times10^{-2}$ \\
$L_{\mathrm{BLR}}$\,[erg\,s$^{-1}$]     & 1.62$\times10^{42}$ & 7.13$\times10^{41}$ & 2.94$\times10^{42}$  & 4.5$\times10^{42}$ \\
$L_{\mathrm{Disk}}$\,[erg\,s$^{-1}$]    & 1.33$\times10^{43}$      & 5$\times10^{42}$ & 2.53$\times10^{43}$   & 4$\times10^{43}$ \\
$L_{\mathrm{Edd}}$\,[erg\,s$^{-1}$]     & 7.8$\times10^{46}$    & 7.8$\times10^{46}$  & 1.95$\times10^{45}$  & 1.95$\times10^{45}$ \\
$\lambda_{\mathrm{Edd,Disk}}$    & 1.7$\times10^{-4}$    & 6.4$\times10^{-5}$  & 0.012  &  0.02 \\
$\lambda_{\mathrm{Edd,BLR}}$     & 2.07$\times10^{-5}$   & 9.1$\times10^{-6}$ & 0.001 &  0.002 \\
$\dot{m}$                               & 0.002   & 0.004  & 0.16    & 0.25  \\ \hline
 \end{tabular*}
\end{table}

\subsubsection{Jet power}

The jet power, computed from SED modeling, regardless of the BH mass, exhibits variations depending on the jet orientation and the size of the emitting region (see Table \ref{Tabl:params2004}). When the jet is oriented at $\theta=2^{\circ}$, with dissipation occurring close to the BH and with a strong magnetic field, we found that $P_r\approx P_B$ for the model with $M_\mathrm{BH}=6\times10^8\,M_\odot$, and $P_B > P_r$ when $M_\mathrm{BH}=1.5\times10^7\,M_\odot$.

On the other hand, for the models with a larger $\theta$, the dissipation region is located at larger distances, well beyond the DT. As a consequence, the magnetic field is weaker, resulting in a lower magnetic jet power $P_r$>$P_B$.

In all models, the power carried by electrons and cold protons dominates the total jet power. This dominance implies a highly efficient conversion of magnetic energy into kinetic energy of the particles, with a part of the magnetic energy preserved as Poynting flux. Interestingly, this is a characteristic observed in AGNs fueled by inefficient accretion modes or ADAFs. In such cases, these sources predominantly emit their energy in kinetic form through radio jets powered via the Bondi mechanism \citep[e.g.,][it is worth mentioning that they consider sub-Eddington accreting SMBHs and low excitation radio galaxies (LERGs) sources]{1995ApJ...452..710N,2007MNRAS.381..589M,2007MNRAS.376.1849H}. These results suggest that the presence of an inefficient disk in this source is plausible. Furthermore, ADAFs are also observed to be associated with jets that display evidence of interaction with the large-scale environment \citep[][]{2016A&A...588A.146S}. The emission observed in the radio maps of this source, shows a potential interaction between the jet and the surrounding medium, a characteristic observed in some CSS sources \citep[e.g.,][]{1998PASP..110..493O,2021A&ARv..29....3O}. Consequently, we can infer that the low accretion mode scenario may point to the hybrid nature of the source.

In Figure \ref{fig:2004-447_power} we show the posterior distribution of the jet power components derived from the MCMC fit, compared to values reported in the literature for a sample of $\gamma-$NLS1, $\gamma-$CSS, and FSRQ (for the MCMC plot see Appendix \ref{fig:2004-447_mcmc}).

\subsection{3C 286}

This source is one of the few CSS detected in $\gamma-$rays up to date, and one of the 5 listed in the 4FGL \cite{2020ApJS..247...33A}. \cite{1982MNRAS.198..843P} categorized it for the first time as a CSS, and later, \cite{2017FrASS...4....8B} identified it as a NLS1 due to its optical spectrum features. As a CSS, it exhibits a compact morphology typical of the class, with a core-jet structure visible on pc scales. Based on its apparent proper-motion speed measured from radio maps, an upper limit of $48^{\circ}$ for the inclination angle of its jet has been estimated \citep[][and references therein]{2017MNRAS.466..952A}. The jet structure observed in the radio maps corresponds to a projected linear size of 910\,pc, and around 600\,pc the jet is seen to bend \citep{2017MNRAS.466..952A}. Therefore, it is crucial to consider the scale at which the angle of inclination of the jet is estimated since it could bend at different scales.
Employing a one-zone model, we cannot reproduce both radio emission and gamma-rays at angles close to 48$^{\circ}$, primarily due to the dependence of inverse Compton emission on the jet orientation angle \citep[see e.g.,][]{2016ApJ...830...94F}. From our results, in Figure \ref{fig:minuit3c286}, we can observe a significant difference in the radio band between the two models. With a viewing angle of $\theta=3^{\circ}$ and the emitting region at $\approx1.6$\,pc, we observe self-absorption of the synchrotron emission around 10$^{11}$\,Hz, with the lower frequencies not accurately reproduced. Meanwhile, with $\theta\sim7^{\circ}$ and the emitting region at $\approx7.19$\,pc, the model effectively reproduces all radio emissions, providing a better description of the entire SED. The bulk Lorentz factor in both models was found to be 11.39 and 10.36, larger than those of $\gamma$-CSS \citep[between 2.4 and 5.5, ][]{2020ApJ...899....2Z} and more consistent with values of $\gamma$-NLS1 $\langle 10.8\pm2.4 \rangle$. The low values of bulk Lorentz factor reported in \cite{2020ApJ...899....2Z} are associated with larger viewing angles. It is important to mention that \cite{2020ApJ...899....2Z} employ a two-zone model, estimating larger viewing angles ($\theta=12.7^{\circ}$ for 3C\,286). However, they still differ from estimates based on radio maps. This suggests that the gamma-ray emission necessarily originates from a region of the jet oriented at a relatively small viewing angle.

\subsubsection{Jet power}
\label{3C286Jetpower}
We have found that in both models the jet power of this source is magnetically dominated, i.e., $P_B>P_r$ and $P_B>P_e$. In Figure \ref{fig:3c286_power}, we present the posterior distribution of the jet power and its different components, derived from the MCMC fit for the CSS scenario (see top panel of Figure \ref{fig:3c286_power}), in comparison with the jet power distribution of a sample of $\gamma-$NLS1, $\gamma$-CSS, and FSRQ \citep[taken from][respectively]{2014Natur.515..376G,Paliya_2019,2020ApJ...899....2Z}. As observed, the derived magnetic power falls within the range of the $\gamma$-CSS \citep[$5.98\times10^{44}-6.91\times10^{46}\mathrm{erg\,s^{-1}}$,][]{2020ApJ...899....2Z}, while the radiation power resembles that of the NLS1 class, falling within the lower end of the distribution \citep[$5.49\times10^{43}-4.78\times10^{46}\mathrm{erg\,s^{-1}}$, see,][]{Paliya_2019}. We notice that a jet highly magnetized was found for the $\gamma$-CSS by \cite{2020ApJ...899....2Z}. Typically, for powerful blazars like FSRQs, where the luminosity is predominantly in the high energy component, interpreted as EC, the $P_B$ is never dominating. \cite{Paliya_2019} have found a similar behavior for the $\gamma$-NLS1s sample, also finding a radiative power, $P_r$ to be more than an order of magnitude lower than that derived for blazars studied in \cite{2017ApJ...851...33P}. With a one-zone model, we have obtained an accretion disk luminosity and a magnetically dominated jet, consistent with the results obtained by \cite{2020ApJ...899....2Z}. 

\subsection{PKS 0440-00}

In this study, we have presented two different approaches for modeling the broadband SED of PKS\,0440-00, to find the best model for reproducing its broadband emission and investigate whether the derived parameters can support the NLS1 classification proposed by \cite{2021Univ....7..372F}.
In the first scenario, we have successfully modeled the SED from radio to gamma-rays using an SSC+EC model with the emitting region located outside the BLR but inside the DT. When considering equation \ref{Rdt} to estimate $R_{\mathrm{DT}}$, even when it implies an oversimplification of the dusty torus component, it provides a good description of the emission by incorporating the composition and size of the dust grains, reproducing the broadband SED with a statistic of $\chi^2=0.86$.

\begin{figure}
    \includegraphics[width=0.9\columnwidth]{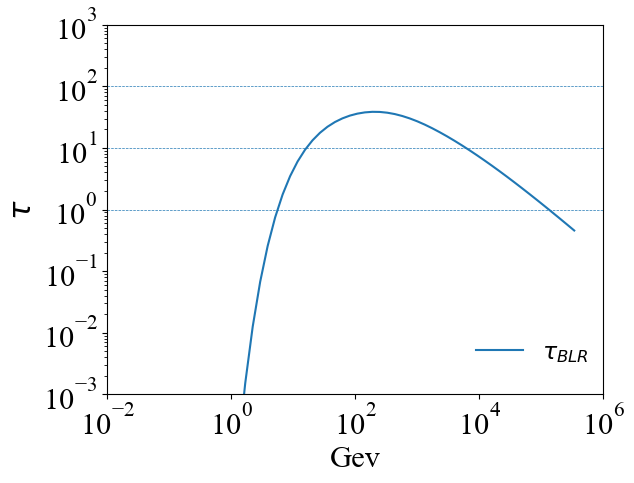}
    \caption{Optical depth as a function of $\gamma$-ray photon energy, for the PKS\,0440-00 model, where the emitting region is located within the BLR radius.}
    \label{fig:tau}
\end{figure}

The second model (see bottom panel of Figure \ref{fig:0440_minuit}), when the dissipation region is located within the BLR, is disfavored due to the tension with radio data under the synchrotron self-absorption frequency. Additionally, there is a significant pair production within the BLR, due to the interaction between photons produced by broad lines and the blob $\gamma-$ray emission \citep[see e.g.,][]{2010ApJ...717L.118P}, producing a significant $\gamma-$ray absorption above 10\,GeV (see Figure \ref{fig:tau}, for the plot of the optical depth for pair production). As a consequence, we conclude that placing the emitting region outside the BLR but inside the DT,  provides a better description of the observed emission.

\subsubsection{PKS 0440-00 FSRQ or NLS1?}
\cite{2021Univ....7..372F} pointed out that the optical spectrum of this source shows an H$\beta$ line with a FWHM $<$ 2000 km\,s$^{-1}$ \citep[][]{2012ApJ...748...49S}, suggesting a NLS1 classification. This finding is relevant since it is well-known that NLS1s and FSRQs present similar characteristics. Particularly, the broadband SED of $\gamma$-NLS1 and FSRQ show: a steep spectrum in gamma-rays, prominent accretion disk emission in the optical-UV range, a Compton-dominated SED, and synchrotron emission peaking at radio to sub-mm wavelengths (indicative of Low Synchrotron Peak sources, LSP). These characteristics suggest a comparable environment around the jet for both types. Consequently, distinguishing the nature of PKS\,0440-447 based on the SED modeling is not straightforward.

We derived a black hole mass of $M_{\mathrm{BH}}=1.73\times10^8\,M_{\odot}$ that falls within the distribution values estimated for $\gamma$-NLS1 \citep[5$\times 10^6$-$7\times 10^8\,M_{\odot}$, ][]{Paliya_2019}. The bulk Lorentz factor of $\Gamma=15.49$, falling within the estimated values for $\gamma$-NLS1 and FSRQ, according to \cite{Paliya_2019} and \cite{2015MNRAS.448.1060G}. We found an accretion disk luminosity of $L_{\mathrm{Disk}}=1.71\times10^{46}\,\mathrm{erg\,s^{-1}}$, falling at the upper end of the values estimated for $\gamma$-NLS1 \citep[$\approx1.69\times10^{46}\,\mathrm{erg\,s^{-1}}$, see][]{Paliya_2019}. FSRQs typically exhibit even higher luminosities, on the order of $10^{47}\,\mathrm{erg\,s^{-1}}$ \citep[see e.g.,][]{2015MNRAS.448.1060G}. One characteristic of $\gamma$-ray emitting NLS1s is the complexity of their X-ray emission, due to the possible contribution of the disk corona (not considered in our models), SSC emission, and EC emission. In our model, we find that the SSC component is crucial to fit the X-ray emission. We found that the jet power is dominated by the bulk motion of matter instead of Poynting flux ($P_\mathrm{e}>P_\mathrm{B}$), and is one order of magnitude larger than the disk luminosity. This composition of the jet power is found in FSRQ sources \citep[see e.g.,][]{2010AIPC.1242...43G}.

Although the fitting-derived parameters, such as those mentioned above, are not decisive in defining the classification of this source, they suggest that based on the modeling of the multifrequency SED, PKS\,0440-00 shares more similarities with the FSRQ class. Therefore, if we consider that the H$\beta$ line is indicative enough for classifying the source as NLS1, we could suggest it as a potential hybrid source between the FSRQ and NLS1 classes.
In Figure \ref{fig:posterior_0440} we show the posterior distribution of the jet power components derived from the MCMC fit, compared to values reported in the literature for a sample of $\gamma-$NLS1, $\gamma-$CSS, and FSRQ (for the MCMC plot see Appendix \ref{fig:0440_mcmc}).

\subsection{Accretion-jet power}

\begin{figure*}
    \centering
	\includegraphics[width=0.95\textwidth]{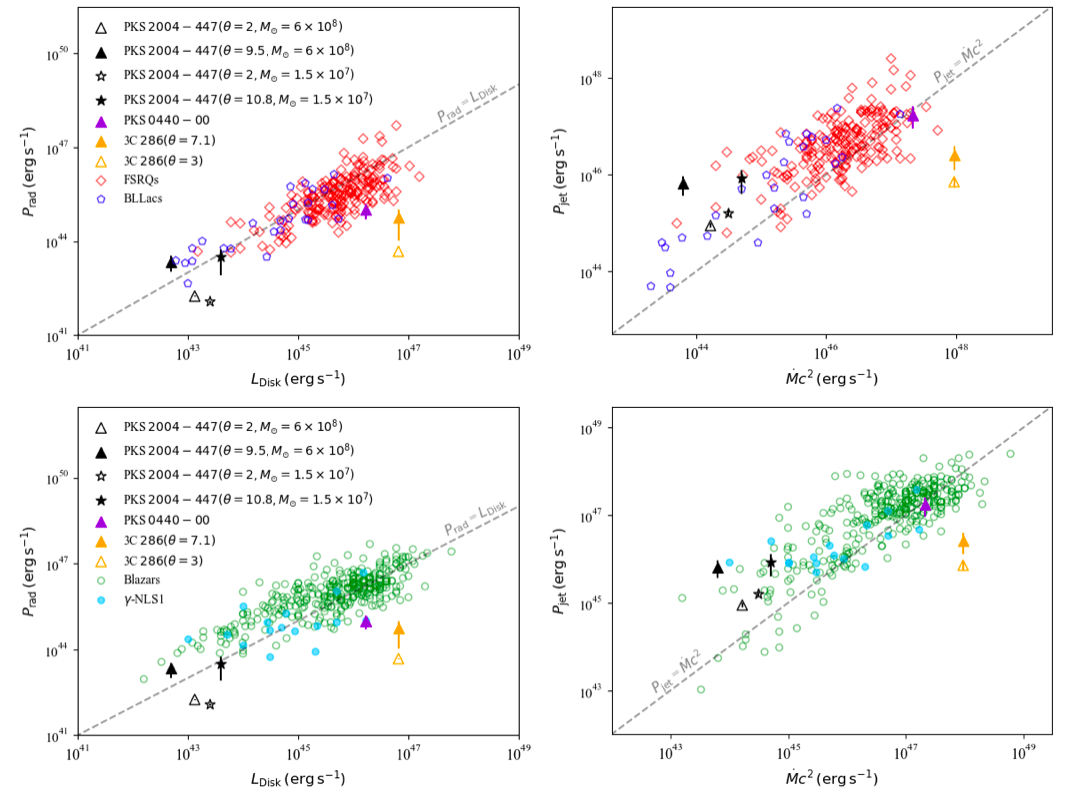}
    \caption{Left: Radiative jet power as a function of the accretion disk luminosity. Right: Total jet power as a function of the total accretion power, $\dot{M}c^2$ (assuming the accretion efficiency reported in the Tables \ref{Tabl:params2004}, \ref{Tabl:params3c286} and \ref{Tabl:params0440}. As reported in the references, $\eta=0.3$ for samples taken from the literature). Different symbols correspond to the values estimated from the different models of the sources. For comparison, in the top panels, we show the FSRQs, BL\,Lacs studied in \citet{2014Natur.515..376G}, while the bottom panels show the so called gamma-loud blazars studied in \citet{2017ApJ...851...33P} and the $\gamma-$NLS1 sample drawn from \citet{Paliya_2019}. The dashed line indicates the equality. }
    \label{Pj_accr}
\end{figure*}

\begin{figure*}
    \centering
	\includegraphics[width=0.9\textwidth]{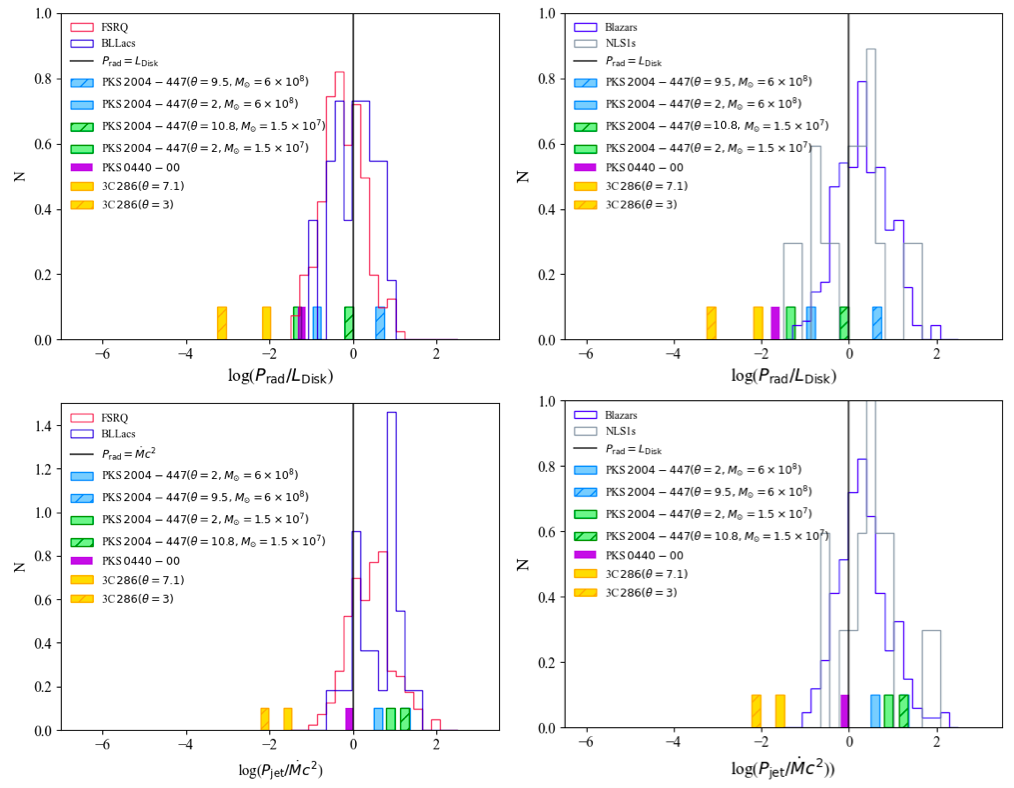}
    \caption{Top: Histograms of the $P_\mathrm{rad}/L_\mathrm{disk}$ ratio. Bottom: Histograms of the $P_\mathrm{jet}/\dot{M}c^2$ rate. Left: FSRQ and BL\,Lacs are shown as a comparison to our sources, drawn from \citet{2014Natur.515..376G}. Right: Blazars and $\gamma-$NLS1 are sourced from \citet{Paliya_2019}.  Bottom panels. The black line corresponds to equality. }
    \label{hist_G}
\end{figure*}

The results of our models can be used to explore the connection between accretion power and the energetic budget of the jet, and the implications in terms of a model-dependent classification. We follow the approach of \cite{2014Natur.515..376G} and, in Figure \ref{Pj_accr}, we present the jet radiative power as a function of the accretion disk luminosity (left panels), and the total jet power as a function of the total accretion power (right panels). The power from our sources is computed as a posterior outcome from the MCMC sampler, hence the bar indicates the 2-$\sigma$ confidence level from our modeling. We stress that this approach allows us to provide a better assessment of the power estimation when compared to large samples, showing an intrinsic dispersion.
The data in the upper panels correspond to our three sources, together with the FSRQs and BL\,Lacs studied by \cite{2014Natur.515..376G}. In the lower panels, we compare our sources with the sample of \cite{2017ApJ...851...33P}, where no distinction is made between FSRQs and BL\,Lacs, and we also include the $\gamma-$NLS1 studied by \cite{Paliya_2019}.  It is well-established in blazars that the jet power is often comparable to or even exceeds the disk luminosity \citep{1991Natur.349..138R}, anyhow, a model-dependent estimate of the jet power, or sample selection criteria, can lead to systematic biases, related to different features/assumptions in the underlying model or in the selected sample. Indeed, in the Ghisellini sample, most sources exhibit a $P_\mathrm{rad}$ lower than $L_\mathrm{Disk}$, in contrast to the Paliya sample, where sources are mainly distributed above the equality line, hence, it is worth comparing our sources to both the samples. Despite the aforementioned difference in $P_\mathrm{rad}$, between the two samples, we notice that our sources occupy, for both the samples, a low radiative-power region, with $P_\mathrm{rad}<L_\mathrm{Disk}$ (see left panels of Figure \ref{Pj_accr}, and top panels of Figure \ref{hist_G}), except for the case of PKS\,2004-447 model with $\theta=9.5^{\circ}$. Interestingly, the distribution of our sources, both in the  $P_\mathrm{rad}$ vs  $L_\mathrm{Disk}$ and $P_\mathrm{jet}$ vs $\dot{M}c^2$ planes, falls within the region of $\gamma$-NLS1s limited by  $P_\mathrm{rad} \lesssim 10^{47} \mathrm{erg\,s^{-1}}$. Even though we notice that 3C\,286 shows a very low $P_\mathrm{rad}$ compared to the $L_\mathrm{Disk}$,  consistently with earlier results obtained by \cite{2020ApJ...899....2Z} for the $\gamma-$CSS.

\section{Conclusion}
\label{sect:conc}
Our analysis of a sample of hybrid sources, based on the multiwavelength SED numerical modeling and an MCMC sampling of the parameter spaces, allowed us to explore different physical parameters, useful in the classification of the sources, such as the jet orientation angle and the connection between the accretion disk and the jet power.

The main results of the multiwavelength analysis of the physical properties of our sources are summarized below.

\begin{itemize}

 \item  We found that  PKS\,2004-447  exhibiting a low Compton-dominated SED along with a relatively low $L_\mathrm{Disk}$ at a large viewing angle, can be modeled as an SSC-dominated object. In the case of an aligned jet with $\theta=2^{\circ}$, the EC emission dominates the IC bump, in agreement with the presence of strong emission lines in the optical band. This might suggest that this source is in a transitional phase, either shifting from an efficient to an inefficient accretion mode or transitioning from a young to a more evolved state. The hybrid nature of this source is evident across different wavelengths.

 \item In the case of 3C\,286, our findings indicate that using a one-zone model limits the viewing angle to $\sim7^{\circ}$, mainly because of its impact on gamma-ray emission. In terms of the disk-jet connection, its behavior appears to resemble that of $\gamma-$NLS1, while the high magnetization of the jet is consistent with values found for $\gamma$-CSS sources, supporting the hybrid nature of the source. We plan a follow-up publication on this source, using more sensitive X-ray data to look for possible spectral signatures of corona emission, and the corresponding impact on the modeling.

 \item PKS\,0440-00 exhibits characteristics common to both NLS1 and FSRQ classes. Our analysis indicates that the best fit is achieved when the emitting region is situated outside the BLR but within the DT. We note the importance of the SSC component in reproducing X-ray emission, as observed in some $\gamma-$NLS1s. However, a comprehensive examination of its optical spectrum is necessary to confirm its classification as NLS1. The presence of similarities with FSRQs and $\gamma-$NLS1 classes suggests the potentially hybrid nature of this source.
\end{itemize}

We plan a follow-up of this analysis, based on time-evolving modeling the presented SEDs, taking into account both the radiative and accelerating processes and the role of the jet adiabatic expansion, to provide a tighter constraint on the underlying physical scenario.

\section*{Acknowledgements}
We thank the referee for insightful comments, which have significantly improved the manuscript.
J.L.C and E.B. acknowledge support from DGAPA-PAPIIT UNAM through grant IN119123 and from CONAHCYT Grant CF-2023-G-100. This research has made use of the NASA/IPAC Extragalactic Database (NED), which is operated by the Jet Propulsion Laboratory, California Institute of Technology, under contract with the National Aeronautics and Space Administration. This paper makes use of the following ALMA data: ADS/JAO.ALMA\#2011.0.00001.CAL. ALMA is a partnership of ESO (representing its member states), NSF (USA) and NINS (Japan), together with NRC (Canada), MOST and ASIAA (Taiwan), and KASI (Republic of Korea), in cooperation with the Republic of Chile. The Joint ALMA Observatory is operated by ESO, AUI/NRAO and NAOJ. This publication makes use of data products from the Wide-field Infrared Survey Explorer, which is a joint project of the University of California, Los Angeles, and the Jet Propulsion Laboratory/California Institute of Technology, funded by the National Aeronautics and Space Administration. Part of this work is based on archival data, software or online services provided by the Space Science Data Center - ASI.

\section*{Data Availability}
The multiwavelength archival data were taken from \url{https://tools.ssdc.asi.it/SED/}. ALMA data were taken from the public Calibrator Source Catalogue \url{https://almascience.eso.org/sc/}.

\bibliographystyle{mnras}
\bibliography{bibliography} 

\appendix

\section{MCMC fit of the SED models presented for each source}

We present the MCMC fit to the observed data, using parameters derived from the frequentist best-fit model as a starting point. We set flat priors for the MCMC, centered on the best-fit values. For each source, we selected a subset of parameters that significantly influence the understanding of their impact on EC emission and jet orientation. For these parameters, we generated posterior contour maps and posterior distributions. The posterior distributions are plotted with vertical dashed lines indicating the selected quantiles: 0.16, 0.5, and 0.84.
\begin{figure*}
	\includegraphics[width=0.65\textwidth]{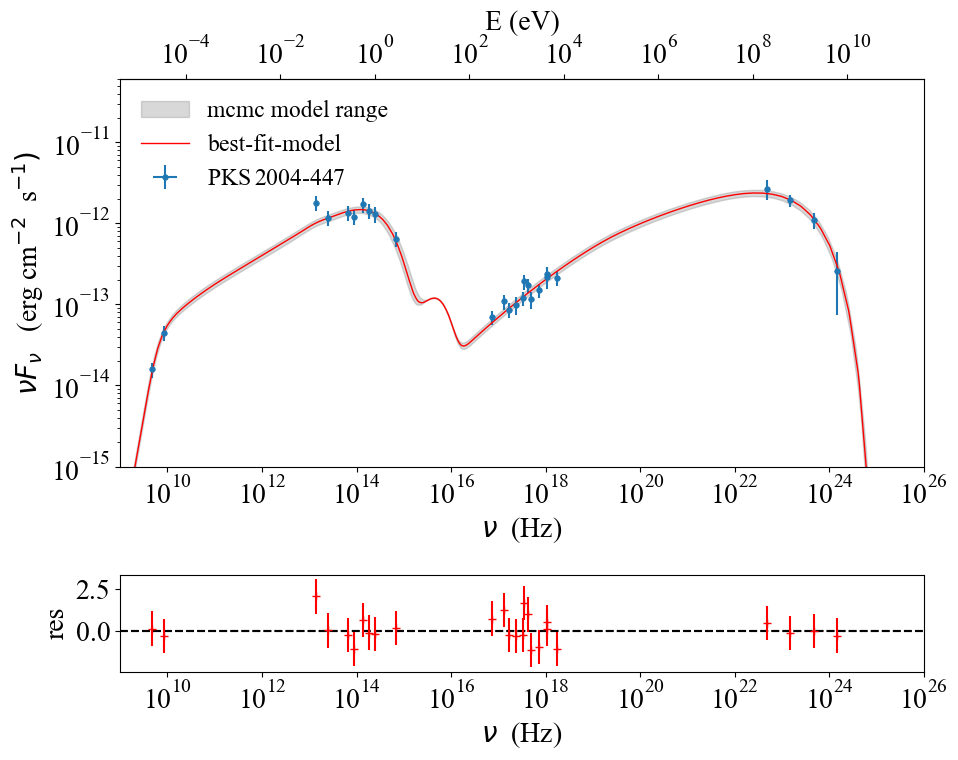}\\
    \includegraphics[width=0.7\textwidth]{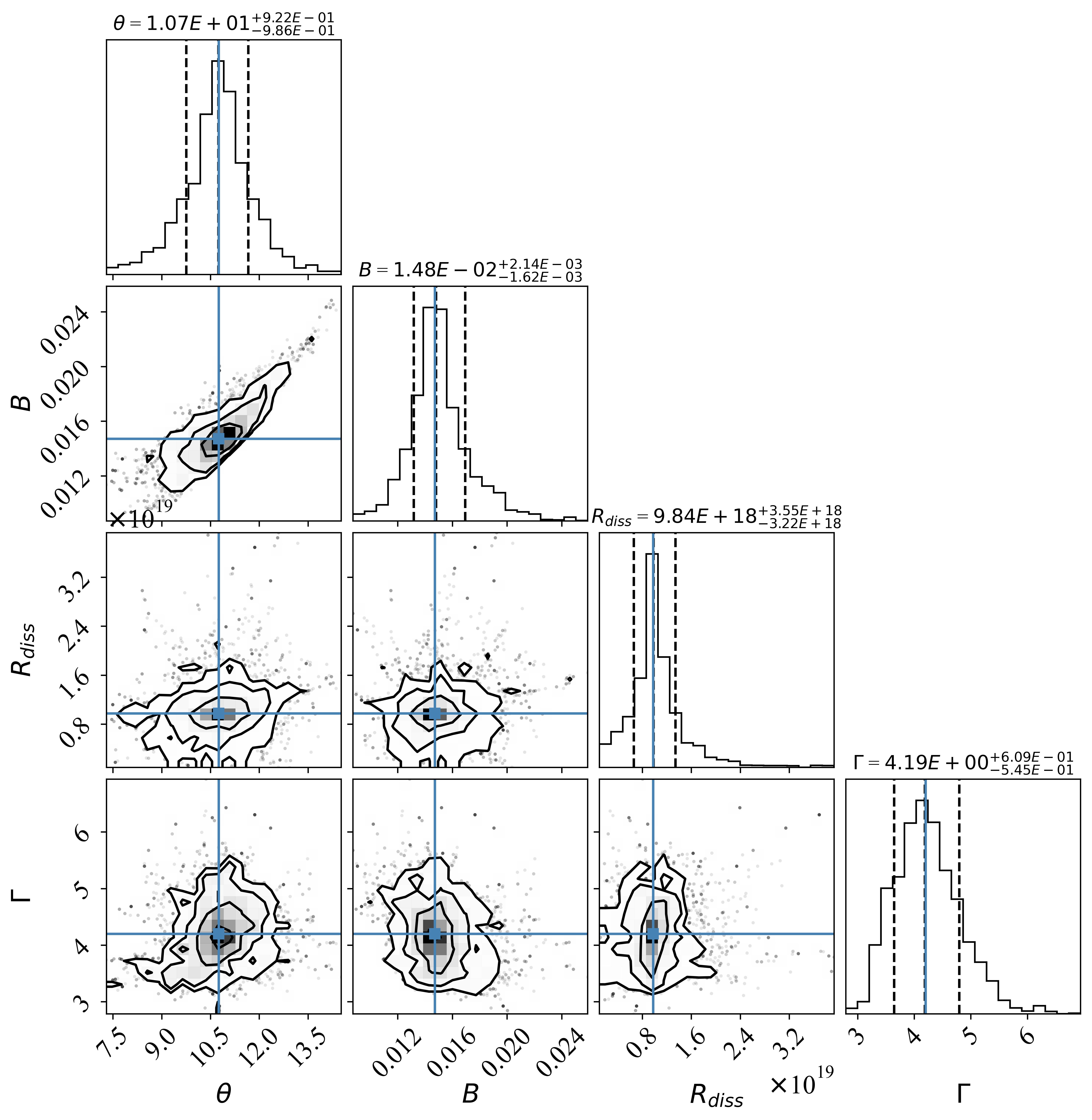}
    \caption{Top panel: MCMC fit to the SED of PKS\,2004-447. The prior distribution is flat and centered on the best-fit values of the model shown in the right bottom panel of Figure \ref{fig:2004-447_MBerton}. Bottom panel: Posterior contour maps illustrating correlations among a sub-sample of parameters. Due to our interest in exploring the impact of the jet orientation and the location of the emitting region, we selected the viewing angle $\theta$, the bulk Lorentz factor, the magnetic field, and the distance from the black hole to the emitting region to generate the posterior contour maps and posterior distribution. There is a strong correlation between the magnetic field ($B$) and the viewing angle ($\theta$).}
    \label{fig:2004-447_mcmc}
\end{figure*}

\begin{figure*}
    \includegraphics[width=0.65\textwidth]{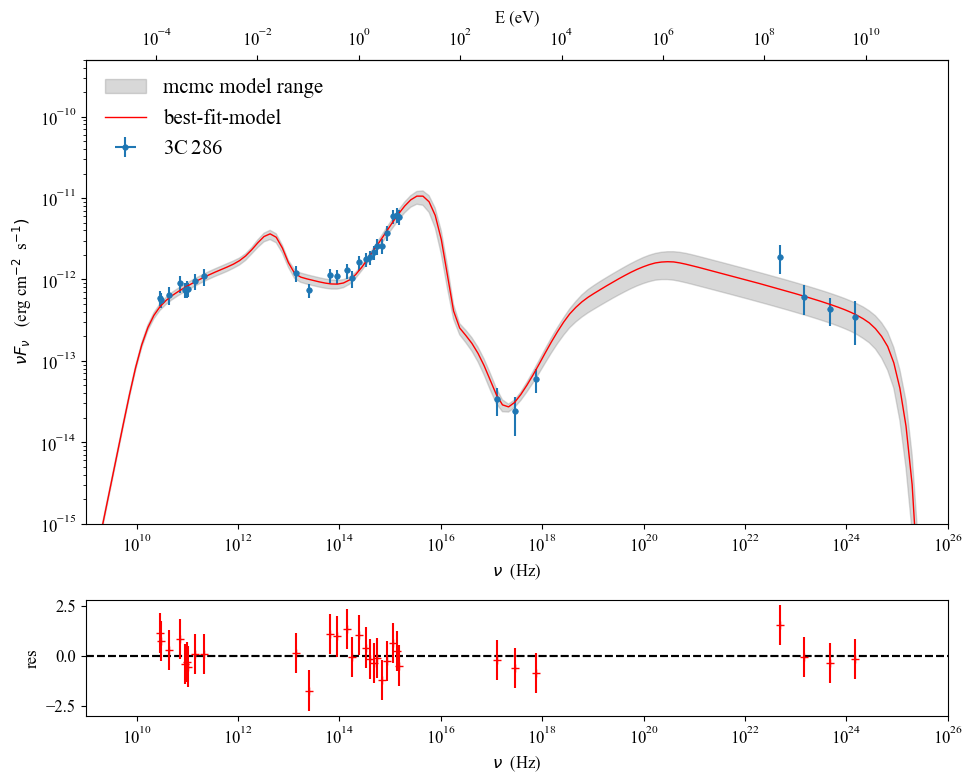}
    \includegraphics[width=0.7\textwidth]{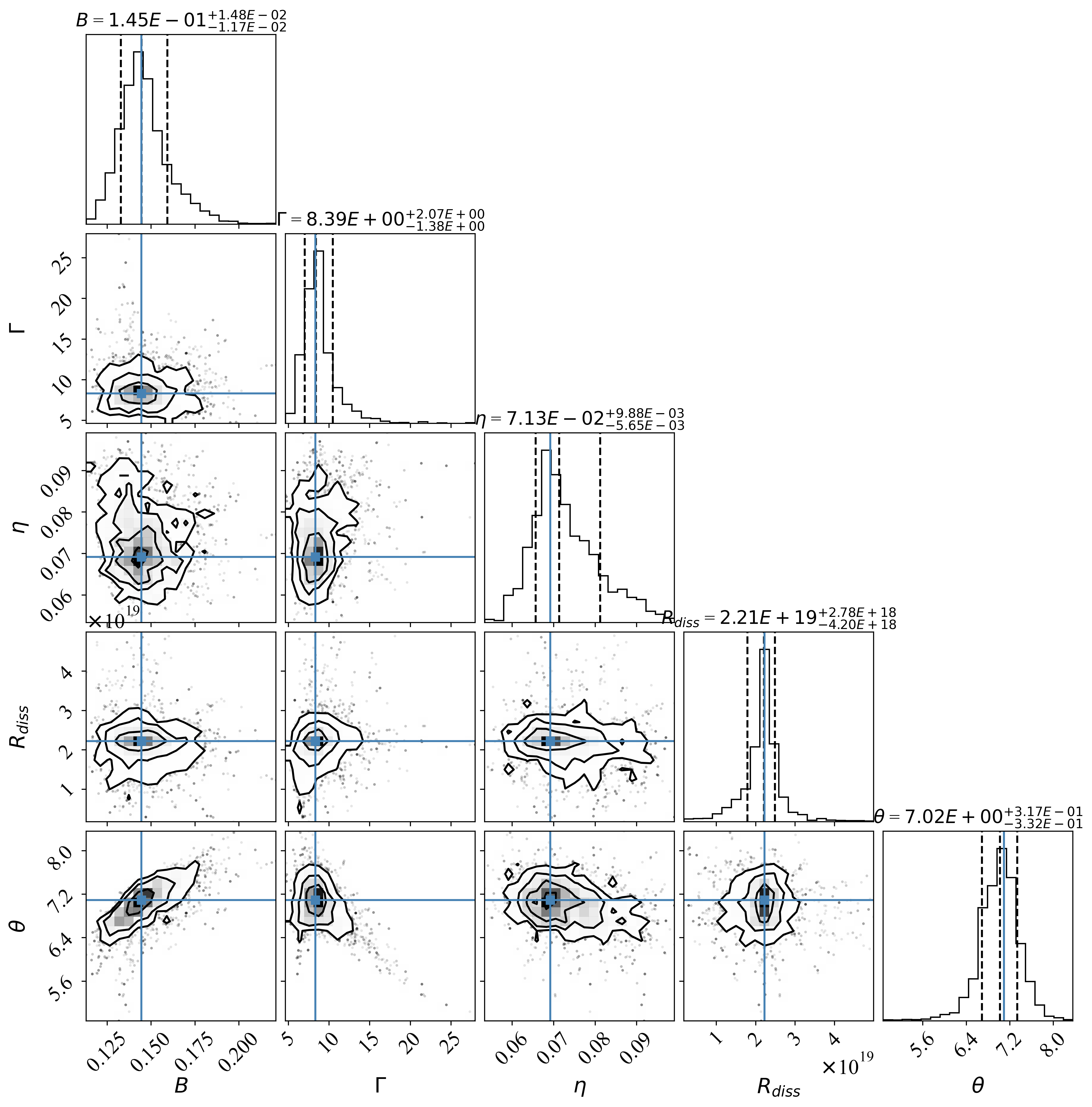}\\
    \caption{Top panel: MCMC fit to the SED of 3C\,286. The prior distribution is flat and centered on the best-fit values of the model shown in the bottom panel if Figure \ref{fig:minuit3c286}. Bottom panel: Posterior contour maps illustrating correlations among a sub-sample of parameters. In order to explore the impact of the jet orientation and accretion disk parameters, we selected the following parameters to generate the posterior contour maps and distribution: the magnetic field, the bulk Lorentz factor, accretion efficiency, the distance from the black hole to the emitting region, and viewing angle. There is an observed correlation between $B$ and $\theta$.}
    \label{fig:3c286_mcmc_angle}
\end{figure*}

\begin{figure*}
    \centering
    \includegraphics[width=0.65
    \textwidth]{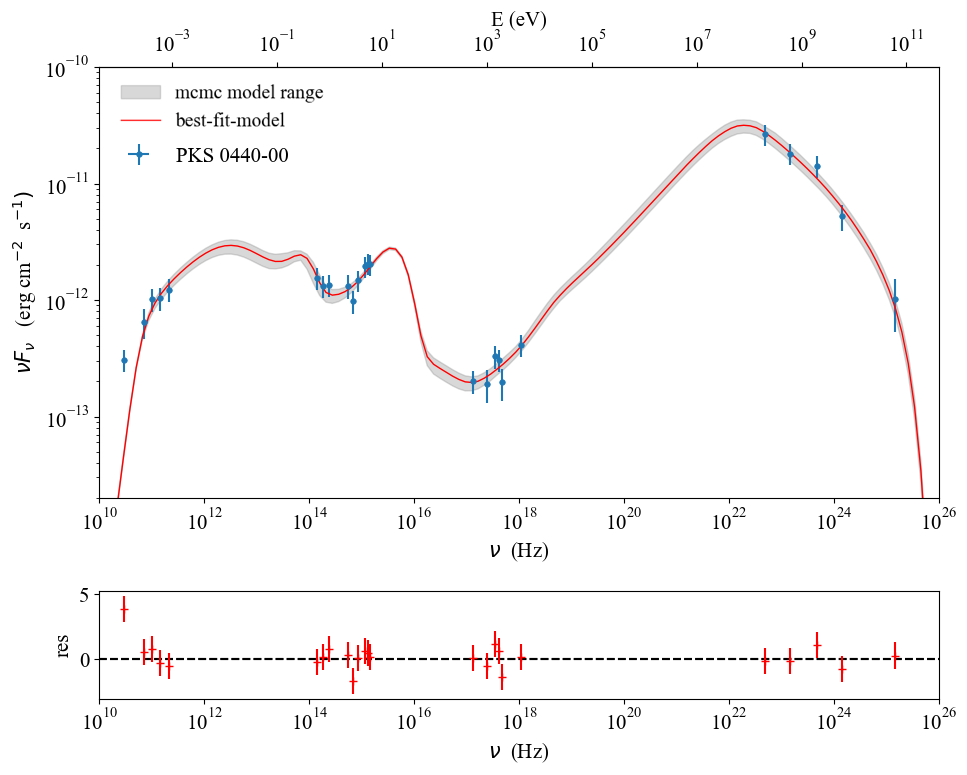}\\
    \includegraphics[width=0.7\textwidth]{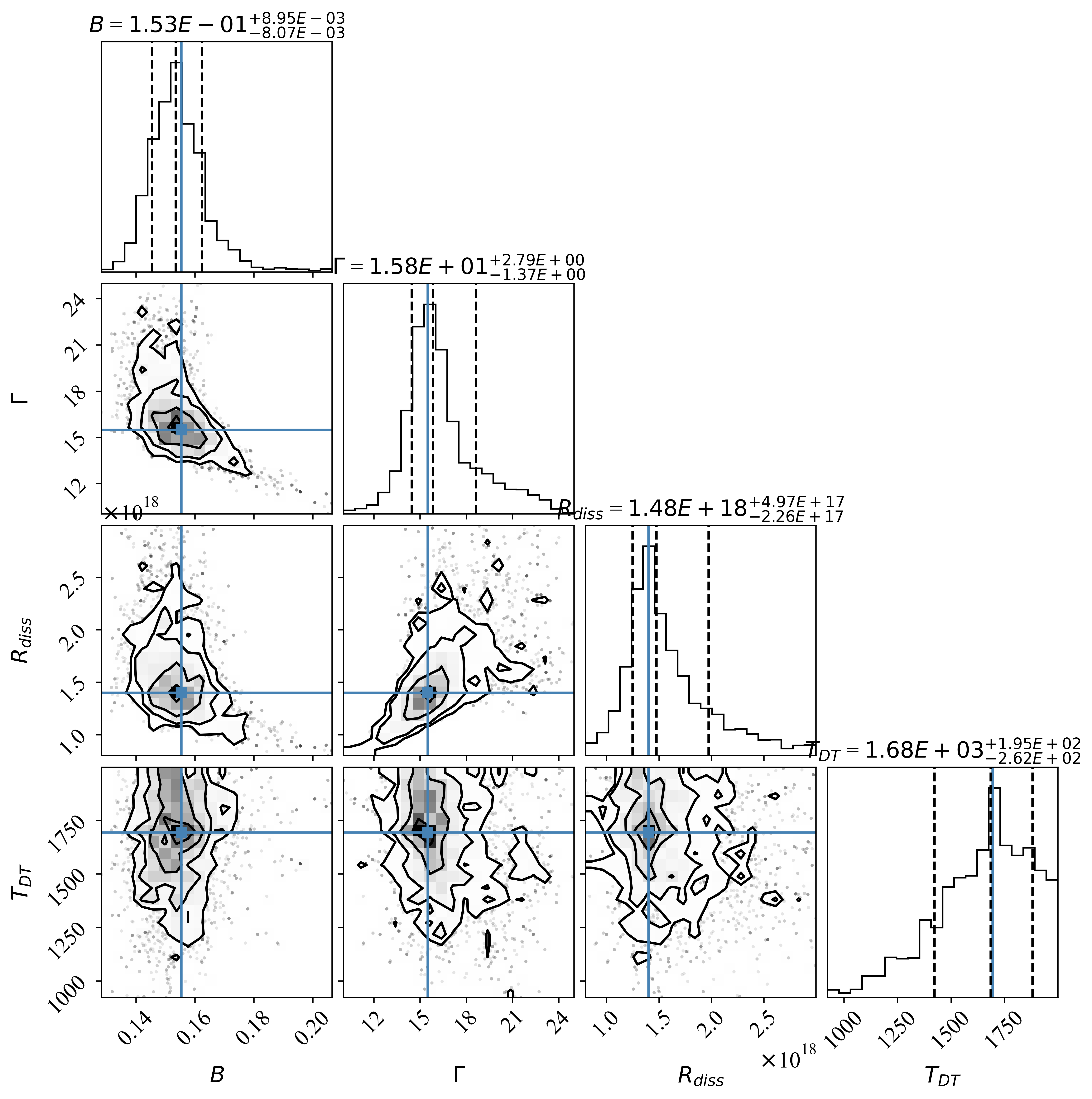}
    \caption{Top panel: MCMC fit to the SED of PKS\,0440-00. The prior distribution is flat and centered on the best-fit values of the model shown in the top panel of Figure \ref{fig:0440_minuit}. Bottom panel: Posterior contour maps illustrating correlations among a sub-sample of parameters. For this source, we generate the posterior contour maps and posterior distribution of the position of the emitting region along the jet, the magnetic field, and the bulk Lorentz factor, as well as the temperature of the dusty torus. There is a observed correlation between the position of the emitting region along the jet ($R_{\mathrm{diss}}$) and the bulk factor ($\Gamma$), and an anticorrelation between the latter and $B$.}
    \label{fig:0440_mcmc}
\end{figure*}

\bsp

\label{lastpage}
\end{document}